\documentclass[preprint]{revtex4}

\usepackage{graphicx}
\usepackage{epsf,epsfig}
\begin{document}
% You should use BibTeX and revtex.bst for references
\bibliographystyle{revtex}

% Use the \preprint command to place your local institutional report
% number  and your conference paper identification number on the
% title page in preprint mode. Multiple \preprint commands are allowed.
\preprint{E3005}
\preprint{NUHEP-EXP/01-053}

%Title of paper
\title{Photon-Photon and Electron-Photon Colliders \\
with Energies Below a TeV}
% Optional argument for running titles on pages
%\title[]{}

% repeat the \author .. \affiliation  etc. as needed
% \email, \thanks, \homepage, \altaffiliation all apply to the current
% author. Explanatory text should go in the []'s, actual e-mail
% address or url should go in the {}'s for \email and \homepage.
% Please use the appropriate macro for the type of information

% \affiliation command applies to all authors since the last
% \affiliation command. The \affiliation command should follow the
% other information

\author{Mayda M. Velasco}
\email[]{Coordinator: mayda.velasco@cern.ch}
%\email[]{schmittm@lotus.phys.nwu.edu}
%\homepage[]{Your web page}
%\thanks{}
%\altaffiliation{}
\affiliation{Northwestern University, Evanston, Illinois 60201, USA}

\author{Michael  Schmitt}
%\email[]{schmittm@lotus.phys.nwu.edu}
%\email[]{mayda.velasco@cern.ch}
%\homepage[]{Your web page}
%\thanks{}
%\altaffiliation{}
\affiliation{Northwestern University, Evanston, Illinois 60201, USA}

\author{Gabriela Barenboim}
%\email[]{gabriela@fnal.gov}
%\homepage[]{Your web page}
%\thanks{}
%\altaffiliation{}
\affiliation{Fermilab, PO Box 500, Batavia, IL 60510-0500, USA}

\author{Heather E. Logan}
%\email[]{logan@fnal.gov}
%\homepage[]{Your web page}
%\thanks{}
%\altaffiliation{}
\affiliation{Fermilab, PO Box 500, Batavia, IL 60510-0500, USA}

\author{David Atwood}
%\email[]{Your e-mail address}
%\homepage[]{Your web page}
%\thanks{}
%\altaffiliation{}
\affiliation{Dept. of Physics and Astronomy, Iowa State University, Ames,
Iowa 50011, USA}

\author{Stephen Godfrey}
%\email[]{Your e-mail address}
%\homepage[]{Your web page}
%\thanks{}
%\altaffiliation{}
\affiliation{Ottawa-Carleton Institute for Physics 
Department of Physics, Carleton University, Ottawa, Canada K1S 5B6}

\author{Pat Kalyniak}
%\email[]{Your e-mail address}
%\homepage[]{Your web page}
%\thanks{}
%\altaffiliation{}
\affiliation{Ottawa-Carleton Institute for Physics 
Department of Physics, Carleton University, Ottawa, Canada K1S 5B6}

\author{Michael A. Doncheski}
%\email[]{Your e-mail address}
%\homepage[]{Your web page}
%\thanks{}
%\altaffiliation{}
\affiliation{Department of Physics, Pennsylvania State University, 
Mont Alto, PA 17237 USA}

\author{Helmut Burkhardt}
%\email[]{Helmut.Burkhardt@cern.ch}
%\homepage[]{Your web page}
%\thanks{}
%\altaffiliation{}
\affiliation{CERN, CH-1211 Geneva 23, Switzerland}

\author{Albert de Roeck}
%\email[]{Albert.de.Roeck@cern.ch}
%\homepage[]{Your web page}
%\thanks{}
%\altaffiliation{}
\affiliation{CERN, CH-1211 Geneva 23, Switzerland}

\author{John Ellis}
%\email[]{John.Ellis@cern.ch}
%\homepage[]{Your web page}
%\thanks{}
%\altaffiliation{}
\affiliation{CERN, CH-1211 Geneva 23, Switzerland}

\author{Daniel Schulte}
%\email[]{Daniel.Schulte@cern.ch}
%\homepage[]{Your web page}
%\thanks{}
%\altaffiliation{}
\affiliation{CERN, CH-1211 Geneva 23, Switzerland}

\author{Frank Zimmermann}
%\email[]{Frank.Zimmermann@cern.ch}
%\homepage[]{Your web page}
%\thanks{}
%\altaffiliation{}
\affiliation{CERN, CH-1211 Geneva 23, Switzerland}

\author{John F. Gunion}
%\email[]{fgucd@higgs.ucdavis.edu}
%\homepage[]{Your web page}
%\thanks{}
%\altaffiliation{}
\affiliation{Davis Institute for High Energy Physics, University of California, Davis, CA 95616, USA}

\author{David M. Asner}
%\email[]{asner1@llnl.gov}
%\homepage[]{Your web page}
%\thanks{}
%\altaffiliation{}
\affiliation{Lawrence Livermore National Laboratory, Livermore, CA 94550, USA}

\author{Jeff B. Gronberg}
\email[]{Coordinator: gronberg1@llnl.gov}
%\homepage[]{Your web page}
%\thanks{}
%\altaffiliation{}
\affiliation{Lawrence Livermore National Laboratory, Livermore, CA 94550, USA}

\author{Tony S. Hill}
%\email[]{hill58@llnl.gov}
%\homepage[]{Your web page}
%\thanks{}
%\altaffiliation{}
\affiliation{Lawrence Livermore National Laboratory, Livermore, CA 94550, USA}

\author{Karl Van Bibber}
%\email[]{e911825@popsicle.llnl.gov}
%\homepage[]{Your web page}
%\thanks{}
%\altaffiliation{}
\affiliation{Lawrence Livermore National Laboratory, Livermore, CA 94550, USA}

\author{JoAnne L. Hewett}
%\email[]{hewett@slac.stanford.edu}
%\homepage[]{Your web page}
%\thanks{}
%\altaffiliation{}
\affiliation{Stanford Linear Accelerator Center, Stanford University, Stanford, California 94309 USA}

\author{Frank J. Petriello}
%\email[]{frankjp@slac.stanford.edu}
%\homepage[]{Your web page}
%\thanks{}
%\altaffiliation{}
\affiliation{Stanford Linear Accelerator Center, 
Stanford University, Stanford, California 94309 USA}

\author{Thomas Rizzo}
%\email[]{frankjp@slac.stanford.edu}
%\homepage[]{Your web page}
%\thanks{}
%\altaffiliation{}
\affiliation{Stanford Linear Accelerator Center, 
Stanford University, Stanford, California 94309 USA}

%Collaboration name if desired (requires use of superscriptaddress
%option in \documentclass). \noaffiliation is required (may also be
%used with the \author command).
%\collaboration{}
%\noaffiliation

\vspace{1cm}
\date{\today}

\vspace{2cm}
\begin{abstract}
We investigate the potential for detecting
and studying Higgs bosons in $\gamma\gamma$   and $e\gamma$ collisions at 
future linear colliders with energies below a TeV. Our study incorporates 
realistic $\gamma\gamma$ spectra based on available laser 
technology, and NLC and CLIC acceleration techniques. Results
include detector simulations. We  study the cases of: a) a SM-like
Higgs boson based on a devoted low energy machine with $\sqrt{s_{ee}}\le 
200$~GeV; 
b) the heavy MSSM Higgs bosons; and c) charged Higgs bosons in  $e\gamma$ 
collisions.
%We also summarizes the utility of the processes $\gamma \gamma \rightarrow
%\gamma \gamma$, $\gamma Z$, and $ZZ$ at a future photon collider  with 
%$\sqrt{s_{ee}}\simeq 1$~TeV in searching
%for new signatures of $CP$ violation in quartic gauge boson self couplings
\end{abstract}
\maketitle

%*****************************************************************
\newcommand{\TEV}{\mbox{TeV}}
\newcommand{\MKM}{\mbox{$\mu$m}}

\def\ggx{{$\gamma\gamma$}\chkspace}
\def\sigg{{$\sigma_{\gamma\gamma}^{tot}$}\chkspace}
\def\fg{{$F^{\gamma}_2$}\chkspace}
\def\xg{{$x_{\gamma}$}\chkspace}
\def\syy{{$\sqrt{s}_{\gamma\gamma}$}\chkspace}
\def\gstar{{$\gamma^*\gamma^*$}\chkspace}
\def\be{\begin{equation}}
\def\ee{\end{equation}}
                         \def\bearr{\begin{eqnarray}}
                         \def\eearr{\end{eqnarray}}
\def\benum{\begin{enumerate}}
\def\eenum{\end{enumerate}}
\def\bitem{\begin{itemize}}
\def\eitem{\end{itemize}}

\newcommand{\ggc}{{\gamma}{\gamma}}          % gg

\newcommand{\pp}{{\rm N}}                  % u quark
\newcommand{\uq}{{\rm u}}                  % u quark
\newcommand{\dq}{{\rm d}}                  % d quark
\newcommand{\sq}{{\rm s}}                  % s quark
\newcommand{\cq}{{\rm c}}                  % c quark
\newcommand{\gq}{{\rm g}}                  % gluon
\newcommand{\cc}{{\rm cc}}                 % cc
\newcommand{\ccb}{{\rm c\bar{c}}}     % ccbar
\newcommand{\bbbar}{{\rm b\bar{b}}}     % ccbar
\newcommand{\ttbar}{{\rm t\bar{t}}}     % ccbar
\newcommand{\WpWm}{{\rm W^+W^-}}     % ccbar
\newcommand{\Acc}{A_{\gamma\pp}^{\ccb}}     % A ypcc
\newcommand{\G}{g}                         % g

\def\beq{\begin{equation}}
\def\eeq{\end{equation}}
\def\bea{\begin{eqnarray}}
\def\beaa{\begin{eqnarray*}}
\def\eea{\end{eqnarray}}
\def\eeaa{\end{eqnarray*}}
\def\gappeq{\mathrel{\rlap {\raise.5ex\hbox{$>$}}
{\lower.5ex\hbox{$\sim$}}}}

\def\lappeq{\mathrel{\rlap{\raise.5ex\hbox{$<$}}
{\lower.5ex\hbox{$\sim$}}}}
\def\bigP{\mbox{\boldmath$P$}}
\def\bigR{\mbox{\boldmath$R$}}
\parskip 0.3cm
% slashes, and d, Re used in math formulas
    \newcommand{\dotg}{\!\!\!/}     \newcommand{\pslash}{p_{\!}\!\!/}
    \newcommand{\dd}{{\rm d}}       \newcommand{\td}{\!{\rm d}}
     \newcommand{\nslash}{n_{\!}\!\!/}                               

%-----------Logan
\def\lsim{\mathrel{\raise.3ex\hbox{$<$\kern-.75em\lower1ex\hbox{$\sim$}}}}
\def\gsim{\mathrel{\raise.3ex\hbox{$>$\kern-.75em\lower1ex\hbox{$\sim$}}}}

%------------Barenboim
\newcommand{\dis}{\displaystyle}
\newcommand{\mathbold}[1]{\mbox{\rm\bf #1}}
\newcommand{\mrm}[1]{\mbox{\rm #1}}
\newcommand{\N}{{\cal N}}
\newcommand{\bla}{\hspace{1cm}}
\newcommand{\nn}{\nonumber}
\newcommand{\eq}[1]{eq.~(\ref{#1})}
\newcommand{\rfn}[1]{(\ref{#1})}
\newcommand{\Eq}[1]{Eq.~(\ref{#1})}
\newcommand{\ep}{\epsilon_K}
\newcommand{\D}{\Delta}

%-----------Gunion
\def\zth{z_{\theta^*}}
\def\gamhtot{\Gamma_{\h}^{\rm tot}}
\def\sig{\sigma}
\def\hhat{\what h}
\def\mhhat{m_{\hhat}}
\def\eq#1{eq.~(\ref{#1})}
\def\Eq#1{Eq.~(\ref{#1})}
\def\fig#1{fig.~\ref{#1}}
\def\Fig#1{Fig.~\ref{#1}}
\def\gamres{\Gamma_{\rm res}}
\def\call{{\cal L}}
\def\cala{{\cal A}}
\def\anti{\overline}
\def\gam{\gamma}
\def\br{B}
\def\ifmath#1{\relax\ifmmode #1\else $#1$\fi}

\def\half{\ifmath{{\textstyle{1 \over 2}}}}
\def\threehalf{\ifmath{{\textstyle{3 \over 2}}}}
\def\quarter{\ifmath{{\textstyle{1 \over 4}}}}
\def\sixth{\ifmath{{\textstyle{1 \over 6}}}}
\def\third{\ifmath{{\textstyle{1 \over 3}}}}
\def\twothirds{{\textstyle{2 \over 3}}}
\def\fivethirds{{\textstyle{5 \over 3}}}
\def\fourth{\ifmath{{\textstyle{1\over 4}}}}
\def\square{\boxxit{0.4pt}{\fillboxx{7pt}{7pt}}\hskip-0.4pt}
    \def\boxxit#1#2{\vbox{\hrule height #1 \hbox {\vrule width #1
             \vbox{#2}\vrule width #1 }\hrule height #1 } }
    \def\fillboxx#1#2{\hbox to #1{\vbox to #2{\vfil}\hfil}    }

\def\ibid{{\it ibid.}}
\def\hf{\hfill}
\def\ie{{\it i.e.}}
\def\etal{{\it et al.}}

\def\cw{c_W}
\def\sw{s_W}
\def\sb{s_\beta}
\def\cb{c_\beta}
\def\dmax{D_{\rm max}}
\def\qks{Q_{KS}}
\def\vqks{\vev{\qks}}
\def\sqks{\sigma_{KS}}
\def\epipr{E_\pi^*}
\def\dedx{dE/dx}
\def\ejet{E_{\rm jet}}
\def\thetamuid{\theta(\mu\mbox{id})}
\def\mrecoil{M_{\rm recoil}}
\def\sigp{\sigma_{\rm P}^{\rm ann}}
\def\signp{\sigma_{\rm NP}^{\rm ann}}
\def\alsp{\alpha_s^{\rm P}}
\def\alsnp{\alpha_s^{\rm NP}}
\def\mpi{m_{\pi}}
\def\sigann{\sigma^{\rm ann}}
\def\vev#1{\langle #1 \rangle}

\def\Eq#1{Eq.~(\ref{#1})}
\def\Ref#1{Ref.~\cite{#1}}

\def\sur{{\wt u_R}}
\def\msur{{m_{\sur}}}
\def\stl{{\wt t_L}}
\def\str{{\wt t_R}}
\def\mstl{m_{\stl}}
\def\mstr{m_{\str}}
\def\sbl{{\wt b_L}}
\def\sbr{{\wt b_R}}
\def\msbl{m_{\sbl}}
\def\msbr{m_{\sbr}}
\def\sq{\wt q}
\def\sqbar{\ov{\sq}}
\def\msq{m_{\sq}}

\def\sel{\wt e}
\def\selbar{\ov{\sel}}
\def\msel{m_{\sel}}
\def\sell{\wt e_L}
\def\msell{m_{\sell}}
\def\selr{\wt e_R}
\def\mselr{m_{\selr}}

\def\cptwo{\wt \chi^+_2}
\def\cmtwo{\wt \chi^-_2}
\def\cpmtwo{\wt \chi^{\pm}_2}
\def\mcptwo{m_{\cptwo}}
\def\mcpmtwo{m_{\cpmtwo}}
\def\stautwo{\wt \tau_2}
\def\mstautwo{m_{\stauone}}

\def\dmchi{\Delta m_{\tilde\chi}}

\def\mth{m_{3/2}}
\def\delgs{\delta_{GS}}
\def\kpr{K^\prime} 

\def\caln{{\cal N}}
\def\cald{{\cal D}}
\def\DM{D$^-$}
\def\DP{D$^+$}

\def\twoloop{two-loop/RGE-improved}
\def\Twoloop{Two-loop/RGE-improved}

\def\mhi{m_{h_1^0}}
\def\etmiss{/ \hskip-9pt E_T}
\def\emiss{~{/ \hskip-8pt E}}
\def\etmin{/ \hskip-9pt E_T^{\rm min}}
\def\etjet{E_T^{\rm jet}}
\def\ptmiss{/ \hskip-9pt p_T}
\def\mslash{~{/ \hskip-9pt M}}
\def\pslash{~{/ \hskip-6pt p}}
\def\ptslash{~{/ \hskip-6pt p}_T}
\def\mchichi{M_{\chi\chi}}
\def\rslash{/ \hskip-9pt R}
\def\susyslash{\susy\hskip-24pt/\hskip19pt}
\def\mmissl{M_{miss-\ell}}
\def\mhalf{m_{1/2}}
\def\aeta{|\eta|}

\def\etc{{\it etc.}}
\def\leff{L_{\rm eff}}
\def\sign{{\rm sign}}

\def\chisq{\chi^2}
\def\cale{{\cal E}}
\def\calo{{\cal O}}
\def\eg{{\it e.g.}}
\def\mhalf{m_{1/2}}

\def\stop{\wt t}
\def\stopone{\wt t_1}
\def\stoptwo{\wt t_2}
\def\mstop{m_{\stop}}
\def\msquark{m_{\wt q}}
\def\mstopone{m_{\stopone}}
\def\mstoptwo{m_{\stoptwo}}

\def\sbot{\wt b}
\def\sbotone{\wt b_1}
\def\sbottwo{\wt b_2}
\def\msbot{m_{\sbot}}
\def\msbotone{m_{\sbotone}}
\def\msbottwo{m_{\sbottwo}}

\def\slep{\wt \ell}
\def\slepbar{\ov{\slep}}
\def\mslep{m_{\slep}}
\def\slepl{\wt \ell_L}
\def\mslepl{m_{\slepl}}
\def\slepr{\wt \ell_R}
\def\mslepr{m_{\slepr}}

\def\To{\Rightarrow}
\def\msusy{m_{\rm SUSY}}
\def\msusyslash{m_{\susyslash}}
\def\susy{{\rm SUSY}}

\def\gl{\wt g}
\def\mgl{m_{\gl}}

\def\tanb{\tan\beta}
\def\cotb{\cot\beta}
\def\mt{m_t}
\def\mb{m_b}
\def\mz{m_Z}
\def\mw{m_W}
\def\mgut{M_U}
\def\mx{M_X}
\def\mstring{M_S}
\def\wp{W^+}
\def\wm{W^-}
\def\wpm{W^{\pm}}
\def\wmp{W^{\mp}}
\def\chitil{\wt\chi}
\def\cnone{\wt\chi^0_1}
\def\cnonestar{\wt\chi_1^{0\star}}
\def\cntwo{\wt\chi^0_2}
\def\cnthree{\wt\chi^0_3}
\def\cnfour{\wt\chi^0_4}
\def\snu{\wt\nu}
\def\snul{\wt\nu_L}
\def\msnul{m_{\snul}}

\def\snue{\wt\nu_e}
\def\snuel{\wt\nu_{e\,L}}
\def\msnuel{m_{\snul}}

\def\snubar{\ov{\snu}}
\def\msnu{m_{\snu}}
\def\mcnone{m_{\cnone}}
\def\mcntwo{m_{\cntwo}}
\def\mcnthree{m_{\cnthree}}
\def\mcnfour{m_{\cnfour}}
\def\h{h}
\def\mh{m_{\h}}
\def\wt{\widetilde}
\def\wh{\widehat}
\def\cpone{\wt \chi^+_1}
\def\cmone{\wt \chi^-_1}
\def\cpmone{\wt \chi^{\pm}_1}
\def\cmpone{\wt \chi^{\mp}_1}
\def\mcpone{m_{\cpone}}
\def\mcpmone{m_{\cpmone}}

\def\cptwo{\wt \chi^+_2}
\def\cmtwo{\wt \chi^-_2}
\def\cpmtwo{\wt \chi^{\pm}_2}
\def\mcptwo{m_{\cptwo}}
\def\mcpmtwo{m_{\cpmtwo}}

\def\staur{\wt \tau_R}
\def\staul{\wt \tau_L}
\def\stau{\wt \tau}
\def\mstau{m_{\stau}}
\def\mstaur{m_{\staur}}
\def\stauone{\wt \tau_1}
\def\mstauone{m_{\stauone}}
\def\sigdmmbar{\overline\sigma_{\dmm}}
\def\gamdmm{\Gamma_{\dmm}}
\def\ep{e^+}
\def\mup{\mu^+}
\def\mum{\mu^-}
\def\taup{\tau^+}
\def\taum{\tau^-}
\def\wpm{W^{\pm}}
\def\hpm{H^{\pm}}
\def\mhm{m_{\hm}}
\def\call{{\cal L}}
\def\calm{{\cal M}}
\def\wtil{\widetilde}
\def\what{\widehat}

\def\ltot{L_{\rm tot}}
\def\taup{\tau^+}
\def\taum{\tau^-}
\def\lam{\lambda}
\def\br{BR}
\def\tauptaum{\tau^+\tau^-}
\def\mbb{m_{b\anti b}}
\def\sprime{{s^\prime}}
\def\rtsprime{\sqrt{\sprime}}
\def\shat{{\hat s}}
\def\rtshat{\sqrt{\shat}}
\def\gam{\gamma}
\def\sigrts{\sigma_{\tiny\rts}^{}}
\def\sigrtssq{\sigma_{\tiny\rts}^2}
\def\sigrtsprime{\sigma_{E}}
\def\nsigrts{n_{\sigrts}}
\def\betao{{\beta_0}}
\def\rhoo{{\rho_0}}
\def\sighbar{\overline \sigma_{\h}}
\def\sighlbar{\overline \sigma_{\hl}}
\def\sighhbar{\overline \sigma_{\hh}}
\def\sighabar{\overline \sigma_{\ha}}
\def\anti{\overline}
\def\epem{e^+e^-}
\def\zstar{Z^\star}
\def\wstar{W^\star}
\def\zstarp{Z^{(\star)}}
\def\wstarp{W^{(\star)}}
\def\mupmum{\mu^+\mu^-}
\def\lplm{\ell^+\ell^-}
\def\brwweff{\br_{WW}^{\rm eff}}
\def\brzzeff{\br_{ZZ}^{\rm eff}}
\def\mstar{M^{\star}}
\def\mstarmin{M^{\star\,{\rm min}}}
\def\drts{\Delta\sqrt s}
\def\rts{\sqrt s}
\def\ie{{\it i.e.}}
\def\eg{{\it e.g.}}
\def\eps{\epsilon}
\def\anti{\overline}
\def\wp{W^+}
\def\wm{W^-}
\def\mw{m_W}
\def\mz{m_Z}
\def\h{h}
\def\mh{m_{\h}}
\def\gamh{\Gamma_{\h}^{\rm tot}}
\def\gamsnu{\Gamma_{\snu}^{\rm tot}}
\def\a{a}
\def\ma{m_{\a}}
\def\hsm{h_{SM}}
\def\mhsm{m_{\hsm}}
\def\gamhsm{\Gamma_{\hsm}^{\rm tot}}
\def\hl{h^0}
\def\mhl{m_{\hl}}
\def\gamhl{\Gamma_{\hl}^{\rm tot}}
\def\ha{A^0}
\def\mha{m_{\ha}}
\def\gamha{\Gamma_{\ha}^{\rm tot}}
\def\hh{H^0}
\def\mhh{m_{\hh}}
\def\hpm{H^{\pm}}
\def\mhpm{m_{\hpm}}

\def\cm{~\mbox{cm}}
\def\fbi{~{\rm fb}^{-1}}
\def\fb{~{\rm fb}}
\def\pbi{~{\rm pb}^{-1}}
\def\pb{~{\rm pb}}
\def\abi{~{\rm ab}^{-1}}
\def\mev{~{\rm MeV}}
\def\gev{~{\rm GeV}}
\def\tev{~{\rm TeV}}
\def\stop{\widetilde t}
\def\mstop{m_{\stop}}
\def\mt{m_t}
\def\mb{m_b}
\def\mm{\mu^+\mu^-}
\def\ee{e^+e^-}

%%%%%%%%%%%%%%%%%%%%%%%%%%%%%%%%%%%%%%%%%%%%%%%%%%%%%%%
\def\MPL #1 #2 #3 {{ Mod.~Phys.~Lett.}~{\bf#1} (#3) #2}
\def\NPB #1 #2 #3 {{ Nucl.~Phys.}~{\bf #1} (#3) #2}
\def\PLB #1 #2 #3 {{ Phys.~Lett.}~{\bf #1} (#3) #2}
\def\PR #1 #2 #3 {{ Phys.~Rep.}~{\bf#1} (#3) #2}
\def\PRD #1 #2 #3 {{ Phys.~Rev.}~{\bf #1} (#3) #2}
\def\PRL #1 #2 #3 {{ Phys.~Rev.~Lett.}~{\bf#1} (#3) #2}
\def\RMP #1 #2 #3 {{ Rev.~Mod.~Phys.}~{\bf#1} (#3) #2}
\def\ZPC #1 #2 #3 {{ Z.~Phys.}~{\bf #1} (#3) #2}
\def\IJMP #1 #2 #3 {{ Int.~J.~Mod.~Phys.}~{\bf#1} (#3) #2}
\def\NIM #1 #2 #3 {{ Nucl.~Inst.~and~Meth.}~{\bf#1} {#3} #2}
\def\JHEP #1 #2 #3 {{ JHEP}~{\bf#1} (#3) #2}
%%%%%%%%%%%%%%%%%%%%%%%%%%%%%%%%%%%%%%%%%%%%%%%%%

\newcommand{\nc}{\newcommand}
%\nc{\beq}{\begin{equation}}   \nc{\eeq}{\end{equation}}
%\nc{\bea}{\begin{eqnarray}}   \nc{\eea}{\end{eqnarray}}
%\nc{\baa}{\begin{array}}      \nc{\eaa}{\end{array}}
\nc{\bit}{\begin{itemize}}    \nc{\eit}{\end{itemize}}
\nc{\bed}{\begin{description}}    \nc{\eed}{\end{description}}
%\nc{\ben}{\begin{enumerate}}  \nc{\een}{\end{enumerate}}
%\nc{\bce}{\begin{center}}     \nc{\ece}{\end{center}}
\def\beqa{\begin{eqnarray}}
\def\eeqa{\end{eqnarray}}

%**********************************************************************
% body of paper here - Use proper section commands
% References should be done using the \cite, \ref, and \label commands

\vspace{-1cm}
\section{Introduction}

The option of pursuing frontier physics with real photon beams
is often overlooked, despite many interesting and informative
studies~\cite{gen}.  The high energy physics community has
focussed on charged particle beams for
historical reasons, and risks missing an excellent opportunity
to do exciting physics in the near future, if the $\ggc$ option is ignored.
In the context
of the next generation of accelerators, most people are
comfortable with the idea of colliding TeV electrons and
positrons, but have not really considered the idea of colliding 100~GeV 
photons. Yet this is now feasible, and could deliver crucial and unique
information on the Higgs sector.  For example, $\gamma\gamma$ collisions
offer a unique capability to measure the two-photon width of the Higgs and
to determine its CP composition through control of the photon
polarization. Also, $\gam\gam$ collisions
offer one of the best means for producing a heavy Higgs
boson singly, implying significantly greater mass reach than
$\epem$ production of a pair of Higgs bosons. Our Snowmass working
group~\cite{E3-SO2}
presents a realistic assessment of the prospects for these 
studies based on the current NLC  machine and detector
designs~\cite{nlc_report,Abe:2001gc} for $\sqrt{s_{ee}}$  up to around
600~GeV,  and CLIC-1~\cite{c:clic,cliche} with
$\sqrt{s_{ee}}\simeq 150$~GeV. The expectations 
for TESLA~\cite{TESLA_TDR,teslatdr} can be deduced by multiplying the NLC yields by a
factor of 1.5 to 2, due the larger 
repetition rate and bunch charge.

\section{The machine}
There is great interest in an $e^+e^-$ linear collider, and one is likely
to be built somewhere in the world.  Here we consider 
75~GeV electrons  for NLC and CLIC-1, and 
100 to 350~GeV for NLC.
At and above these energies, all types of machines: NLC/JLC, TESLA and
CLIC, have suitable luminosities for a $\gamma\gamma$ collider.
In all cases, we assume $e^-e^-$ collisions
as our starting point, and the electrons to be  80\% longitudinally polarized.
We prefer to only use electrons, because one can obtain higher luminosity 
and
total $\ggc$ polarization 
than with positrons.

In all cases, a $\ggc$ interaction region would fit into the present 
plans.
Both NLC~\cite{nlc_report} and TESLA~\cite{teslatdr} have plans for
a second, lower energy, interaction
region that can be used for $\ggc$ collisions, while the CLIC-1 based
design that we have developed assumes only one dedicated $\ggc$
interaction region.  We 
refer to the CLIC-1 based design as CLICHE~\cite{cliche},
the ``CLIC Higgs Experiment''.

The photon beams required in a $\ggc$ collider would be produced via the
Compton backscattering of laser light off the high-energy electron beam.
In the electron-laser collision at the conversion point,
the maximum energy of the scattered photons is 
$\omega_m=\frac{x}{x+1}E_0; \;\;
x \approx \frac{4E_0\omega_0}{m^2c^4}
 \simeq 15.3\left[{E_0}/{\TEV}\right]
\left[{\omega_0}/{eV}\right]$, 
where $E_0$ is the electron beam energy and $\omega_0$ the energy of the
laser photon. In connection with NLC studies~\cite{nlc_report}, the case
has been considered of $E_0 =250$~GeV, $\omega_0 =1.17$~eV, i.e., a 
wavelength  of 
$1.054$ \MKM\ , with a 
high power Mercury laser from the LLNL group.
This
would correspond to $x=4.5$ and $\omega_m = 0.82E_0$.

The computation of the luminosity function 
$F(y)=d\call_{\gam\gam}/dy/\call_{\gam\gam}$
\cite{Ginzburg:1983vm,Ginzburg:1984yr}, assuming a short (1-5\,mm)  distance 
from the electron--laser collision to the $\gam\gam$ interaction point, is 
shown in Fig.~\ref{f:gamgamlum_plot} as a function of 
$y=E_{\ggc}\sqrt{s_{ee}^{-1}}$ along with the $\vev{\lam\lam'}$ values,
where the $\lam$'s are the resulting photon beam helicities.
There are three independent choices for $\lam_e$, $\lam'_e$, $P$ and
$P'$, where  $\lam_e=\frac 1 2 P_e$ is the electron helicity and  
$P$ is the laser polarization.
In Fig.~\ref{f:gamgamlum_plot} we give the results for
three independent choices of relative electron
and laser polarization orientations for the values of $x$ relevant
in our studies, $x=5.69$, $x=4.334$ and $x=1.86$.

We observe that  choice (I) of $\lam_e=\lam'_e=0.4$, $P=P'=1$ gives a
large $\vev{\lam\lam'}$ and $F(y)>1$ for small to moderate $y$.
Therefore, it could be interesting with high energy machines seaching
in a broad energy range for $J=0$ particles like heavy Higgs bosons. 
In a machine with 315~GeV electrons and $1.054~\mu$m lasers, for example,
$x=5.69$.  It has been argued in the past that $x>4.8$
is undesirable because it leads to pair creation. However, our studies,
which include these effects, indicate that the resulting backgrounds
are not a problem. 

The choice (II) of $\lam_e=\lam'_e=0.4$, $P=P'=-1$ yields
a peaked spectrum with $\vev{\lam\lam'}>0.85$ at the maximum.
If we use $1.054~\mu$m lasers,  then 
a value of  $x=1.86$ for   103~$\gev$ electrons is obtained, and
we can see that the $\gam\gam$ spectrum peaks at
$E_{\gam\gam}^{peak}\sim 120\gev$. This would be an optimal setting
for  light Higgs boson studies.
If a tripler is added to the laser system, then the wavelength is reduced
by a factor of three, and a $\gam\gam$ spectrum peaking  at 
$E_{\gam\gam}^{peak}=120\gev$
is obtained by operating at $\sqrt{s_{ee}}=160\gev$,
yielding $x=4.334$. The realistic  spectra and luminosities  for such 
cases are plotted in
Fig.~\ref{fig:spectra}, for CLICHE with  $x=4.334$  and NLC with $x=1.86$. 
These results were produced with the CAIN~\cite{cain2} program, which takes into
account the beamstrahlung,  secondary collisions between scattered electrons
and photons from the laser beam and other non-linear effects.
The result is a substantial enhancement of the luminosity in the
low-$E_{\gam\gam}$ region compared to the simple predictions given 
in Fig.~\ref{f:gamgamlum_plot}.  The improvement for $x=4.334$
as compared to $x=1.86$ is clear.   A summary of the expected luminosities
at $E_{\gam\gam}^{peak}$ running with this  polarization configuration 
is shown in Table~\ref{tab:lum}, for different electron beam  energies.
  
Finally, the choice (III) of $\lam_e=\lam'_e=0.4$, $P=1,P'=-1$
gives a broad spectrum, but never achieves large $\vev{\lam\lam'}$.
Large values of $\vev{\lam\lam'}$
are important for suppressing the $c \anti c$ and 
$b\anti b$ continuum 
background to Higgs detection, 
whose leading tree-level term $\propto 1-\vev{\lam\lam'}$, 
making
this configuration not useful for Higgs studies.

\begin{table}[htbp]
\caption{
Luminosities for $J=0$ component of the $\ggc$ energy spectra at
$E_{\ggc}^{peak}$ assuming $\lam_e=\lam'_e=0.4$, $P=P'=-1$, 
choice (II).  Values for  different machines
and beam energies tuned as a Higgs Factory,  
$M_{Higgs}\simeq E_{\ggc}^{peak}$, are given. 
$\vev{\lam\lam'}$  is given at the energy
corresponding to $E_{\ggc}^{peak}$.}
\label{tab:lum}
\begin{center}
\begin{tabular}{lcccc}
\hline
$E_{e}$ (GeV)\,\,\, &\,\,\,${\cal L}_{\ggc}^{peak} (fb^{-1}$/\{bin width GeV\})\,\,\, & $E_{\ggc}^{peak}$ (GeV)\,\,\, &\,\,\, $\vev{\lam\lam'}$ & comment \\
\hline
CLIC--75      &     	4.7 / 3.33 & 115.     & 0.94 & 22000 Higgs/year($10^7$ sec) in SM  \\
\hline
NLC --80      &     	1.7 / 3.33 & 120.     & 0.87 & 11000 Higgs/year($10^7$ sec) in SM    \\
NLC --103 & 		1.5 / 3.33 & 120.     & 0.85 &  9500 Higgs/year($10^7$ sec) in SM \\
NLC --267.5&  	3.4 /11.13 & 406.     & 0.80  & varies, {\it i.e.} see Fig.~\ref{f:sigeff}   \\
NLC --315 & 	3.4 /13.10 & 478.     & 0.79 & varies, {\it i.e.} see Fig.~\ref{f:sigeff} \\
\hline
\end{tabular}
\end{center}
\end{table}

\begin{figure}[h!]
\centerline{\psfig{file=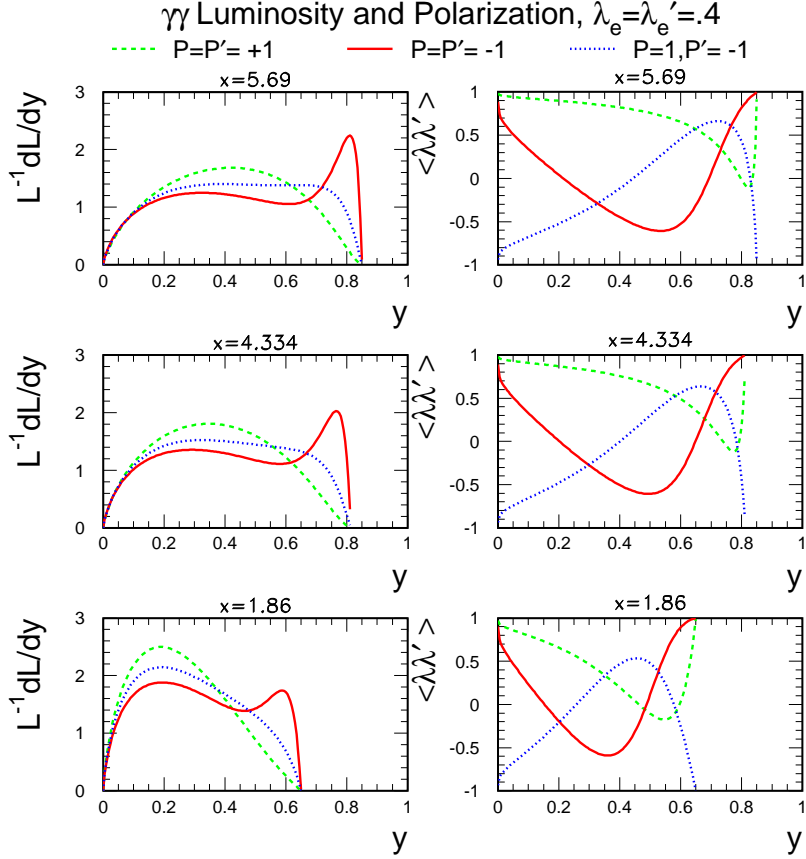,width=12cm}}
\caption[0]{The normalized differential luminosity
  ${1\over \call_{\gam\gam}}{d\call_{\gam\gam}\over dy}$ and the corresponding
  $\protect\vev{\lam\lam'}$ for $\lam_e=\lam'_e=.4$ (80\%
  polarization)
and three different choices of the initial laser photon polarizations
$P$ and $P'$. The distributions shown are for
$\rho^2\ll 1$ \cite{Ginzburg:1983vm,Ginzburg:1984yr}. Results for  $x=5.69$,
$x=4.334$ and $x=1.86$ are compared.
}
\label{f:gamgamlum_plot}
\end{figure}

%%%%%%%%%%%%%%%%%%%%%%%%%%%%%%%%%%%%%%%%%%%%%%%%%%%%%%%%%%%%%%%%%%%
%%%%%%%%%%%%%%%% F I G U R E %%%%%%%%%%%%%%%%%%%%%%%%%%%%%%%%%%%%%%%%%%%%%%%%%%
%%%%%%%%%%%%%%%%%%%%%%%%%%%%%%%%%%%%%%%%%%%%%%%%%%%%%%%%%%%%%%%%%%%%%%%%%%%%%%%
\begin{figure}[tbp]
\begin{center}
\mbox{\epsfig{file=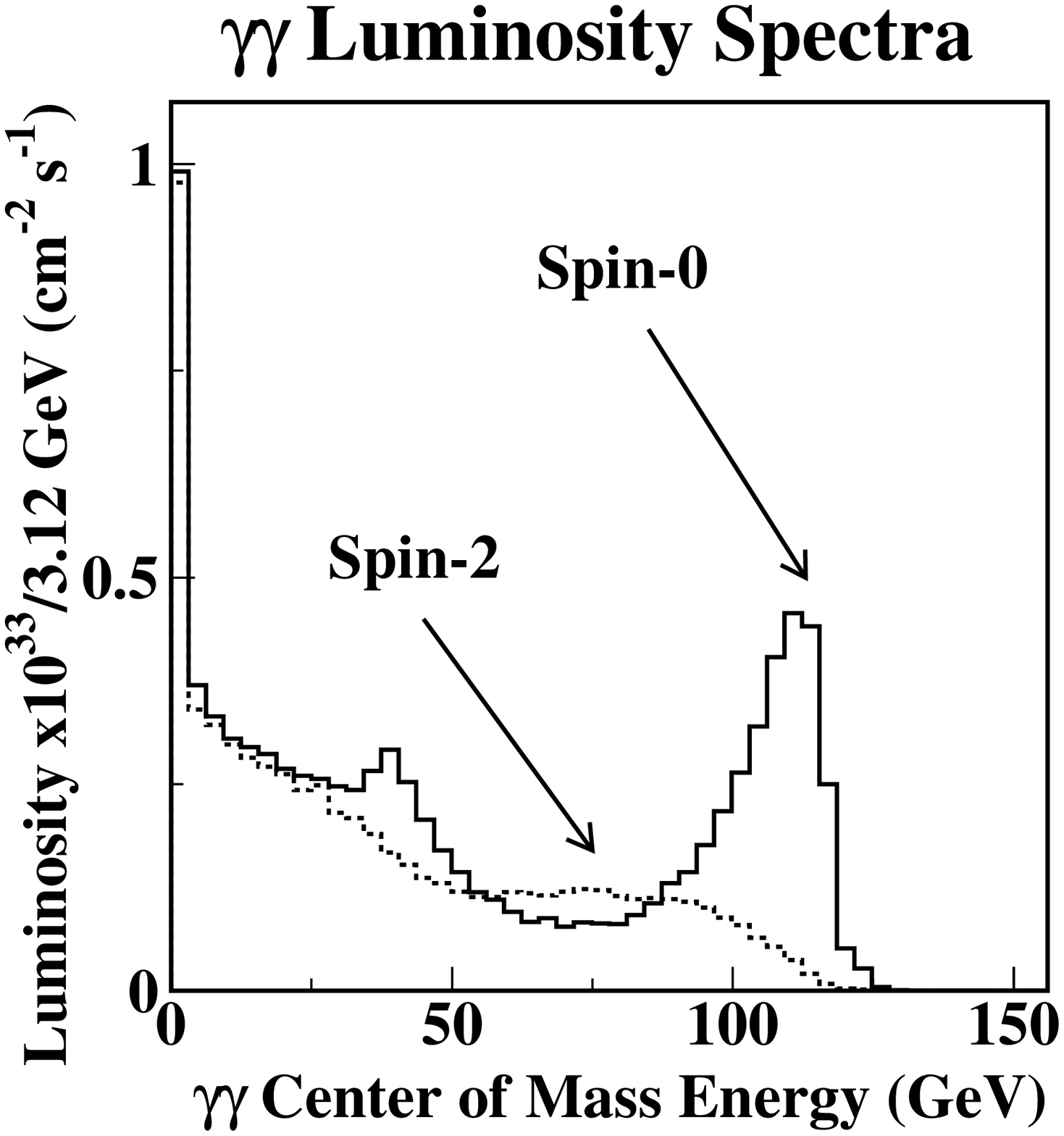,height=5cm}}
\mbox{\epsfig{file=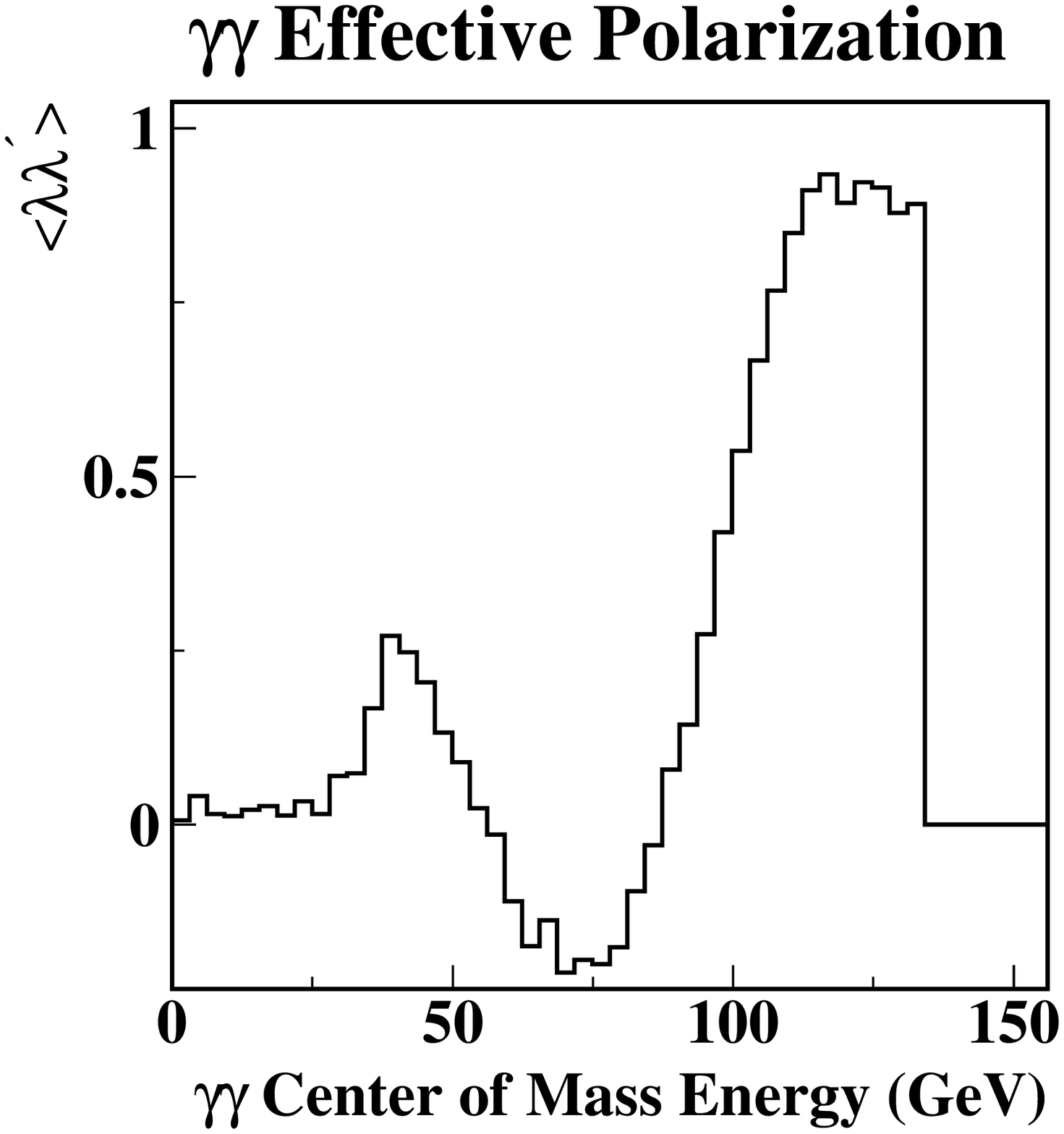,height=5cm}}
\mbox{\epsfig{file=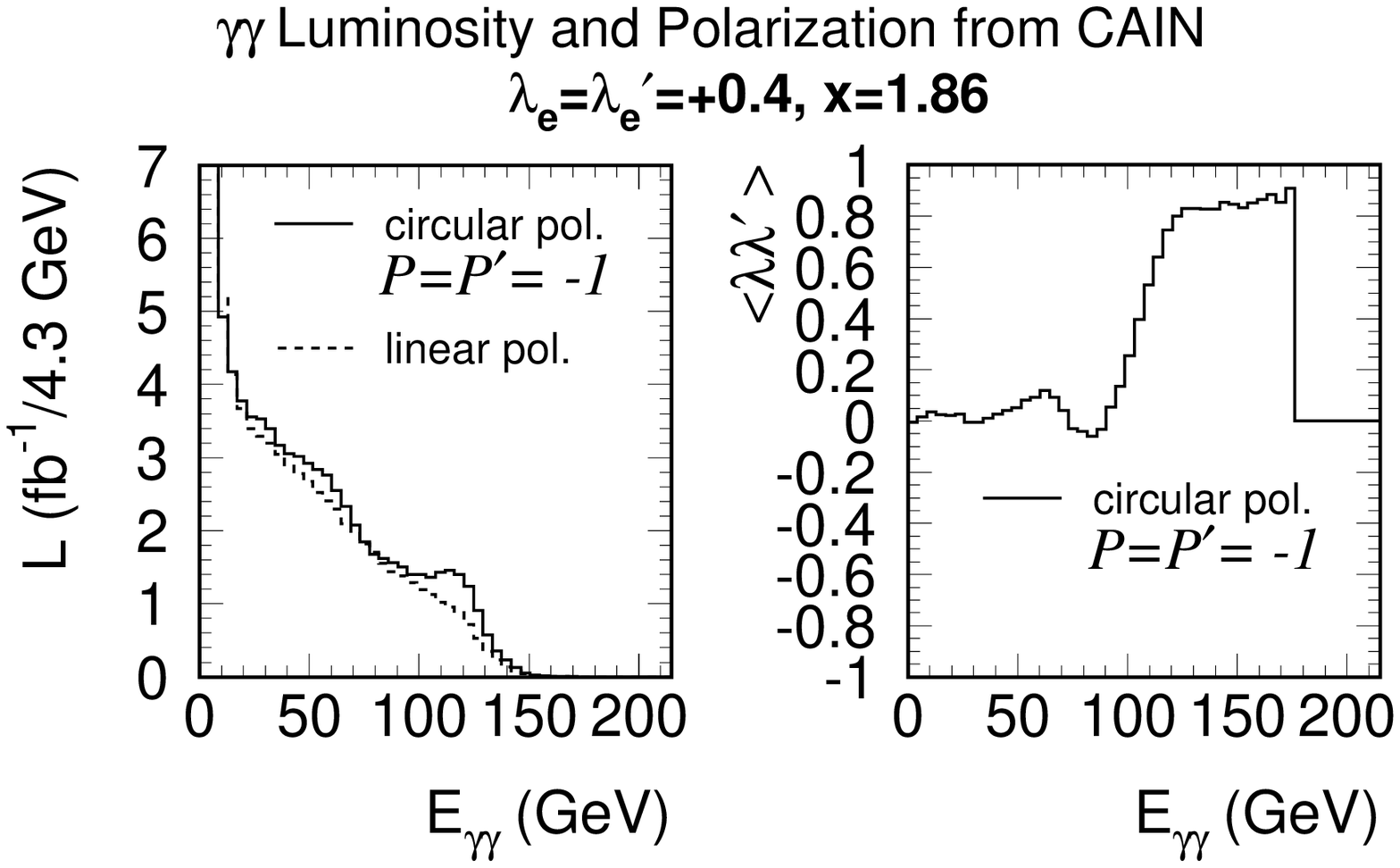,height=7cm}}
\caption[.]{\label{fig:spectra}\it
(a) Luminosity spectra and beam polarization as functions of
${E_{\gamma\gamma}}$ for the CLIC-1 parameters for 75~GeV electrons.
(b) Luminosity spectra and beam polarization as functions of
${E_{\gamma\gamma}}$ for the NLC parameters for 103~GeV electrons.
Also plotted is the corresponding value of $\vev{\lam\lam'}$ 
given by $\vev{\lam\lam'}=(L_{J_z=0}-L_{J_z=2})/(L_{J_z=0}+L_{J_z=2})$.
}
\end{center}
\end{figure}
%%%%%%%%%%%%%%%%%%%%%%%%%%%%%%%%%%%%%%%%%%%%%%%%%%%%%%%%%%%%%%%%%%%

%%%%%%%%%%%%%%%%%%%%%%%%%%%%%%%%%%%%%%%%%%%%%%%%%%%%%%%%%%%%%%%%%%%%%%%%%%%%%
\section{Physics Opportunities --  Higgs Factories}
\label{ana}

All the studies shown below use JETSET fragmentation, the event 
mixture 
predicted by PYTHIA (passed through JETSET) \cite{pythiajetset},
and the LC Fast MC detector simulation within ROOT \cite{Brun:1997pa},
which includes calorimeter smearing and the detector configuration.
The signal is generated using PANDORA plus PYTHIA/JETSET \cite{pandora}.
The luminosity and polarization predictions from the CAIN \cite{cain2}
Monte Carlo  were used to produce the beam spectra. 

\subsection{Light Higgs Measurements}
\label{l_higgs}
\noindent\underline{Mass measurement:}\\
The cross sections for a Higgs boson masses around $115$~GeV as
functions of $E_{CM}(e^- e^-)$ for unpolarized electrons are shown in
Fig.~\ref{fig:excitation}(a). 
We see that the cross section rises rapidly
for $E_{CM}(e^- e^-)$ between 140 and 160~GeV.  
This feature, combined with the large value of  ${\cal B}r(H \to 
b\bar{b})$,
can be used to measure the Higgs mass by sweeping across 
the threshold for Higgs production and measuring how the 
number of $\bar{b}b$ events increases.  Since the position of this threshold 
depends on the Higgs mass, a scan offers the possibility to
measure the Higgs mass kinematically, as developed in~\cite{ohgaki}.
%%%%%%%%%%%%%%%%%%%%%%%%%%%%%%%%%%%%%%%%%%%%%%%%%%%%%%%%%%%%%%%%%%%
%%%%%%%%%%%%%%%% F I G U R E %%%%%%%%%%%%%%%%%%%%%%%%%%%%%%%%%%%%%%%%%%%%%%%%%%
%%%%%%%%%%%%%%%%%%%%%%%%%%%%%%%%%%%%%%%%%%%%%%%%%%%%%%%%%%%%%%%%%%%%%%%%%%%%%%%
\begin{figure}[tbp]
\begin{center}
\resizebox{\textwidth}{!}
{\epsfig{file=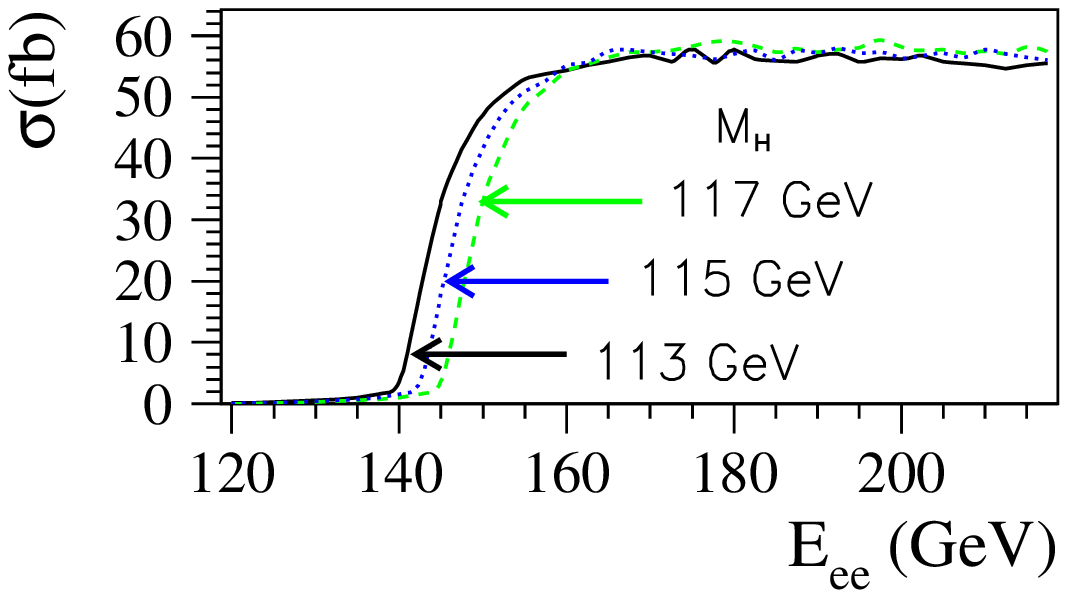,height=6.5cm}
\epsfig{file=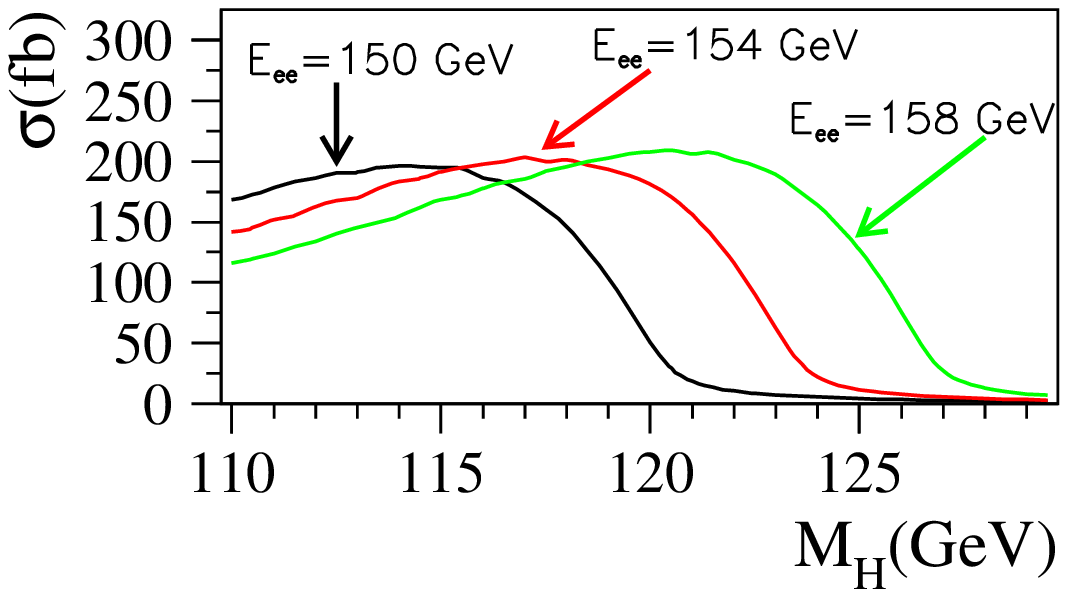,height=6.5cm}}
\caption[.]{\label{fig:excitation}\em
(a) The cross sections for $\gamma \gamma \rightarrow H$ for
different values of $m_H$ as functions of $E_{CM}(e^-e^-)$ 
for unpolarized photons. 
(b) The cross section for $\gamma \gamma \rightarrow H$
as a function of $m_H$ for three different values of  $E_{CM}(e^- e^-)$.
Here the electrons are assumed to be 80\% polarized longitudinally,
and the lasers circularly polarized, so that the produced 
photons are highly circularly polarized at their peak energy.
}
\end{center}
\end{figure}
%%%%%%%%%%%%%%%%%%%%%%%%%%%%%%%%%%%%%%%%%%%%%%%%%%%%%%%%%%%%%%%%%%%

%
%--------------------------------------

We have studied this possibility in the context of $\gamma\gamma$ Higgs
factories, constructed with NLC and CLIC based technologies,
assuming that the Higgs mass is already
known to within a GeV or so, from the Tevatron or the LHC. 
As shown in Fig.~\ref{fig:scan}, there is a point of optimum 
sensitivity to the Higgs mass a few~GeV below the peak of the cross section. 
There is another point  close to the maximum
of the cross section, at which there is no sensitivity to the Higgs mass,
and with maximum sensitivity to $\Gamma_{\gamma\gamma}$, allowing the
separation of these two quantities. These points are illustrated in
Fig.~\ref{fig:scan}.  
%%%%%%%%%%%%%%%%%%%%%%%%%%%%%%%%%%%%%%%%%%%%%%%%%%%%%%%%%%%%%%%%%%%
%%%%%%%%%%%%%%%% F I G U R E %%%%%%%%%%%%%%%%%%%%%%%%%%%%%%%%%%%%%%%%%%%%%%%%%%
%%%%%%%%%%%%%%%%%%%%%%%%%%%%%%%%%%%%%%%%%%%%%%%%%%%%%%%%%%%%%%%%%%%%%%%%%%%%%%%
\begin{figure}[tbp]
\begin{center}
\resizebox{\textwidth}{!}
{\epsfig{file=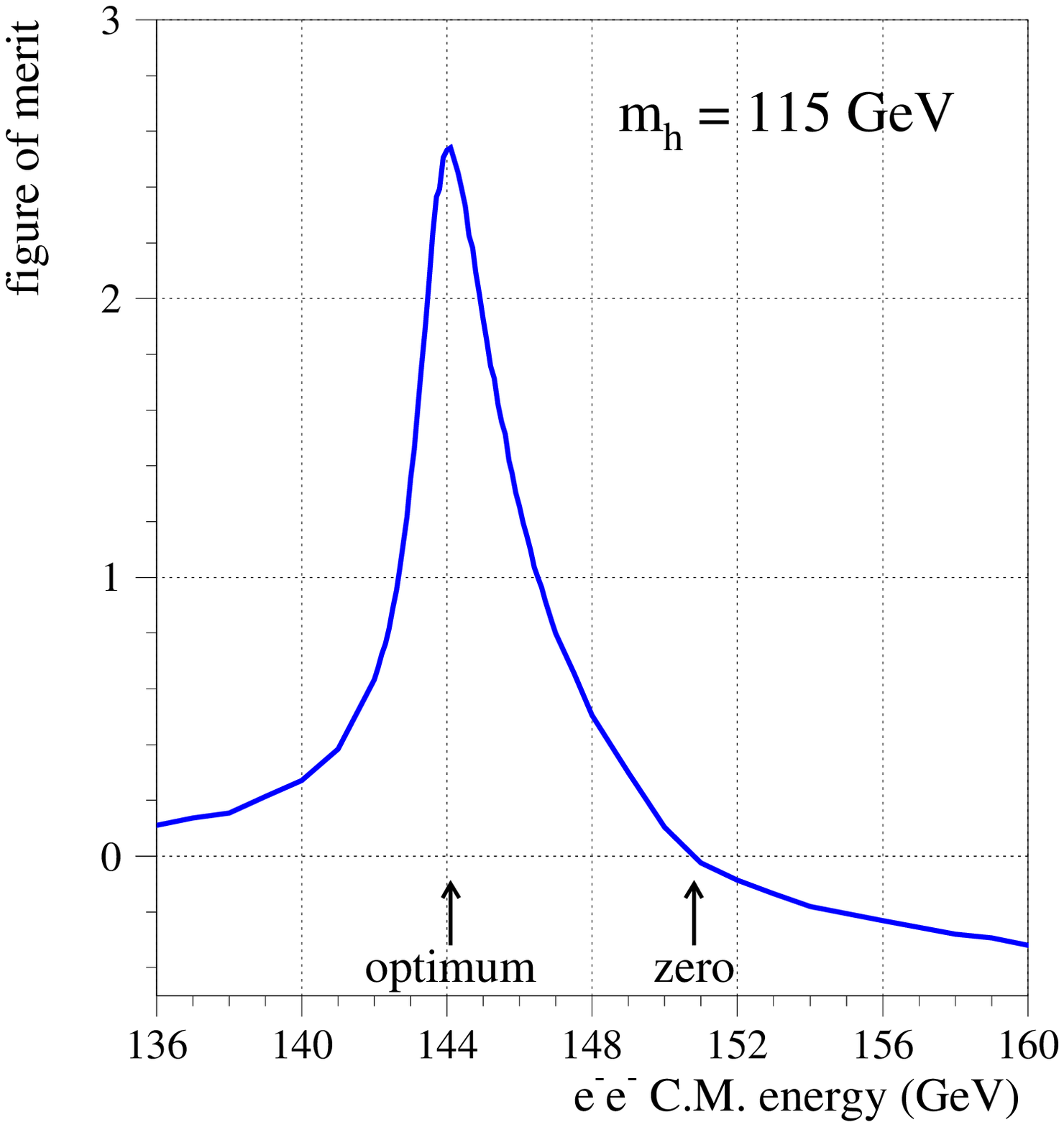,height=7cm}
\epsfig{file=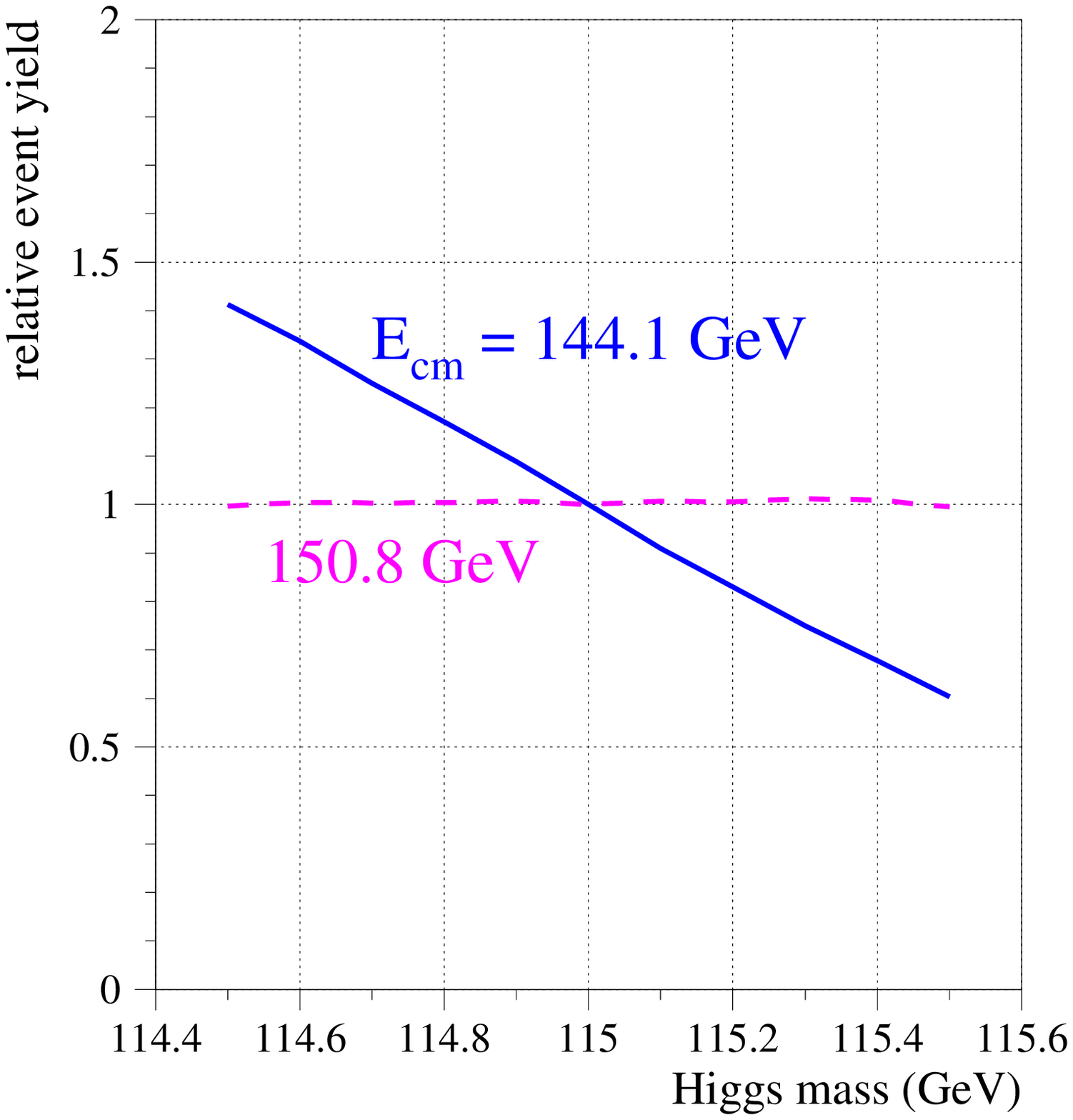,height=7cm}
\epsfig{file=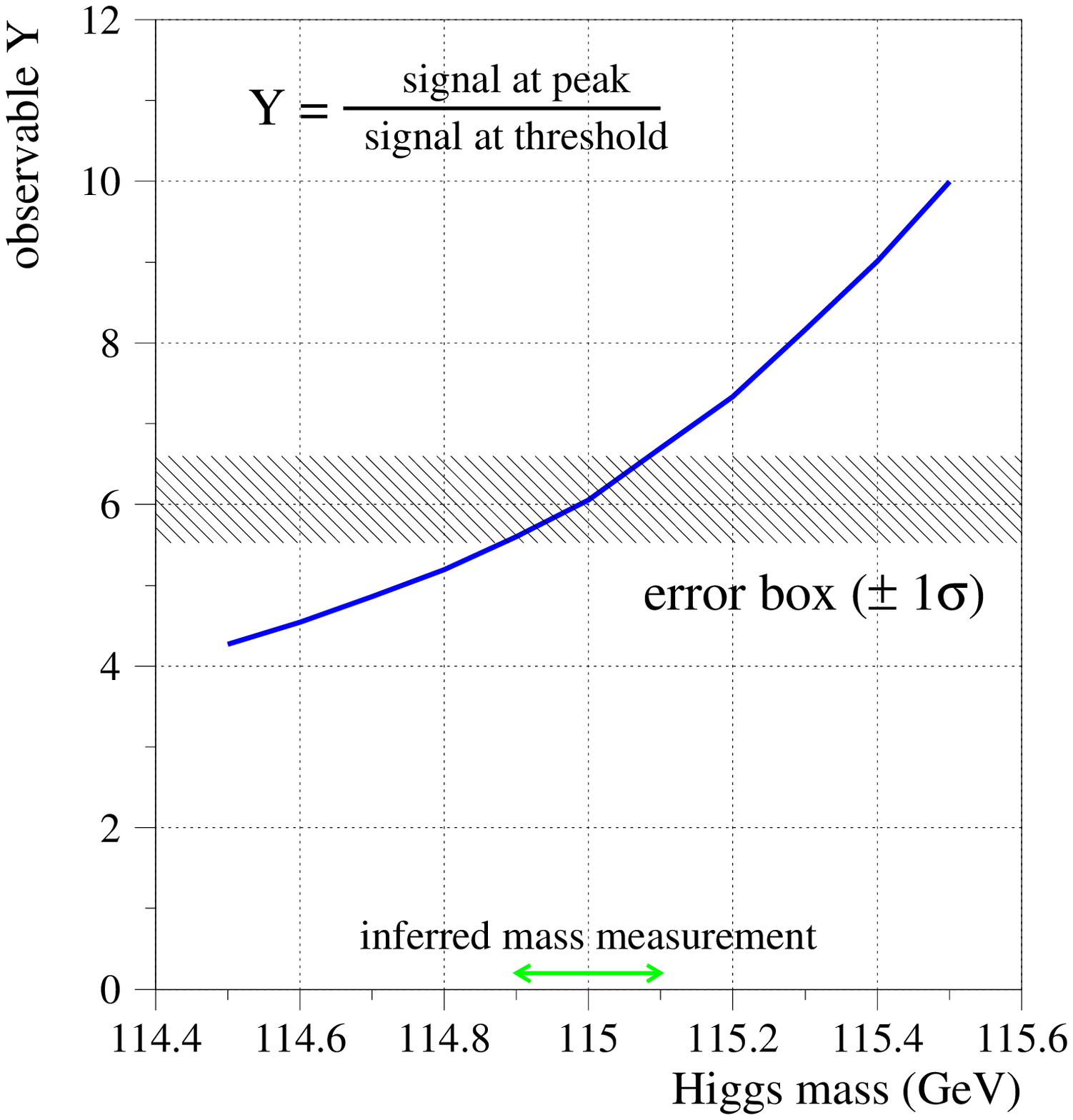,height=7cm}}
\caption[.]{\label{fig:scan}
\em 
(a) A figure of merit quantifying the measurement error on the mass as a function of the $e^-e^-$ center-of-mass energy. The optimum and zero 
sensitivity points are marked.
(b) Relative yield for a 115~GeV Higgs boson at the point of optimum
sensitivity and zero sensitivity to $m_H$.  
(c) Behavior of the observable~$Y  = \frac{N_{\mathrm{peak}} - N_{\mathrm{below}}\cdot r_p} {N_{\mathrm{threshold}} - N_{\mathrm{below}}\cdot r_t}$ 
as a function of $m_H$, and the projected error, where $N$ is the number 
of events in a mass window logged at the peak,
on the threshold, and below threshold, and $r_p$ and $r_t$ are scale
factors to relate the background data taken below threshold to
the expectation at peak and at threshold.
}
\end{center}
\end{figure}
%%%%%%%%%%%%%%%%%%%%%%%%%%%%%%%%%%%%%%%%%%%%%%%%%%%%%%%%%%%%%%%%%%%
This translates into an error on the inferred Higgs mass of~100~MeV. 
More details for our  analysis can be found in~\cite{cliche}.

%--------------------------------------
\noindent\underline{$H \rightarrow {\bar b} b$:}\\
Our analysis includes perturbative QCD backgrounds,
including $\gamma \gamma \rightarrow {\bar b} b(g)$ and
$\gamma \gamma \rightarrow {\bar c} c(g)$.
The ${\bar q} q$ backgrounds are suppressed by choosing like polarizations
for the colliding photons, but this suppression is not so strong when the
final states contain additional gluons.
We assume that there will be a 3.5\%
$c\bar{c}$ contamination and that the $b$ tagging is 70\% efficient for the 
double tag events.  The
final reconstruction efficiency is expected to be 30\%.
More details for our  analysis can be found in~\cite{cliche,GunionHA}.

In the CLICHE design the mass resolution is around 6~GeV with a 
jet energy resolution of $\sigma_E=0.6 \times \sqrt{E}$. The distribution in 
the di-jet invariant mass, $m_{jets}$, for a $m_H=115$~GeV Higgs found in 
this study is shown in Fig.~\ref{fig:hbb}. 
%%%%%%%%%%%%%%%%%%%%%%%%%%%%%%%%%%%%%%%%%%%%%%%%%%%%%%%%%%%%%%%%%%%
%%%%%%%%%%%%%%%% F I G U R E %%%%%%%%%%%%%%%%%%%%%%%%%%%%%%%%%%%%%%%%%%%%%%%%%%
%%%%%%%%%%%%%%%%%%%%%%%%%%%%%%%%%%%%%%%%%%%%%%%%%%%%%%%%%%%%%%%%%%%%%%%%%%%%%%%
\begin{figure}[tbp]
\begin{center}
\mbox{\epsfig{file=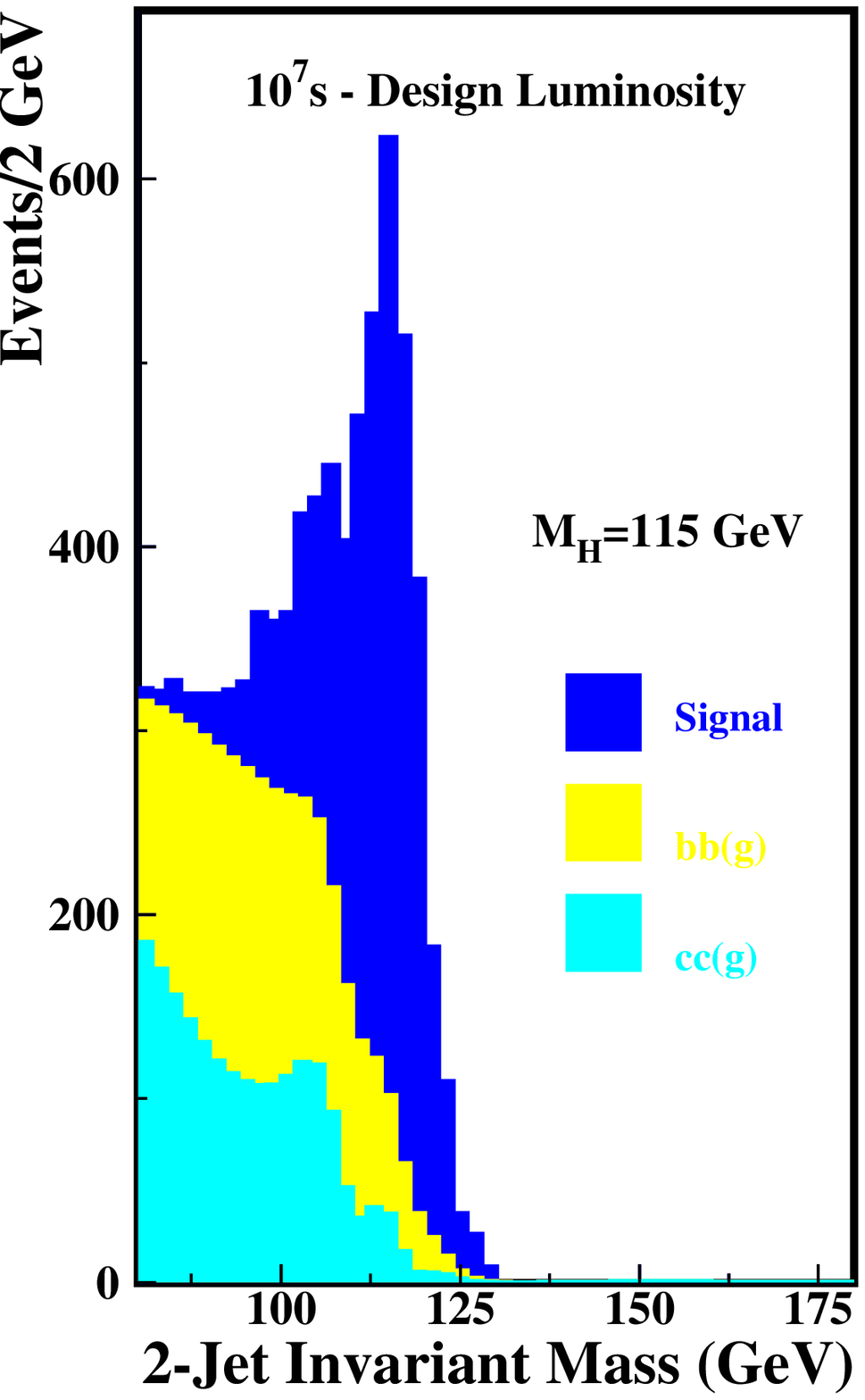,height=8cm}}
\mbox{\epsfig{file=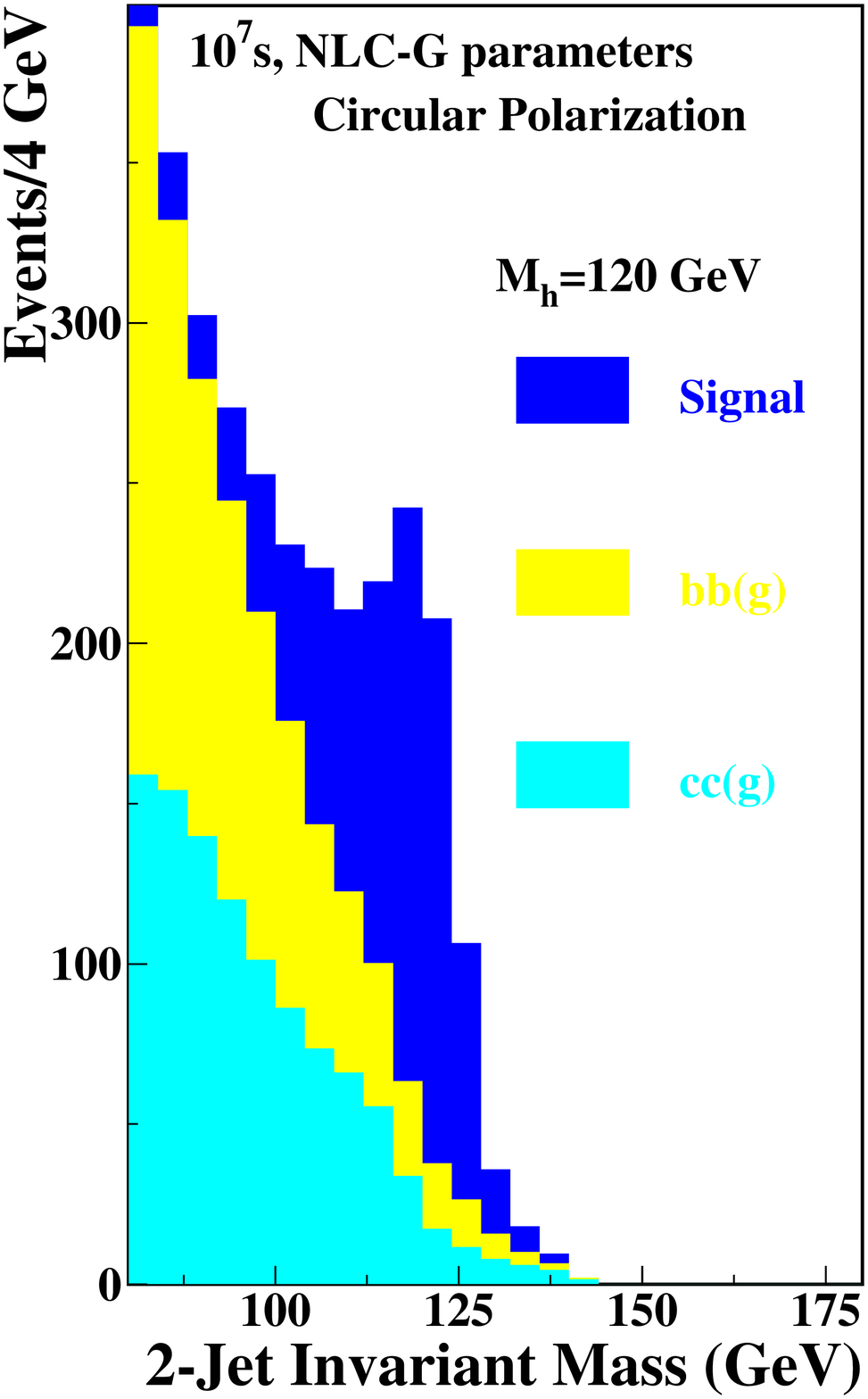,height=8cm}}
\caption[.]{\label{fig:hbb}      
\em
Observability of the $H \rightarrow {\bar b} b$  decay mode 
at   CLICHE with $\sqrt{s_{ee}}
= 150$~GeV~\cite{cliche}, and at NLC with $\sqrt{s_{ee}}
= 206$~\cite{nlc_report}. 
}
\end{center}
\end{figure}
%%%%%%%%%%%%%%%%%%%%%%%%%%%%%%%%%%%%%%%%%%%%%%%%%%%%%%%%%%%%%%%%%%%
A clear signal peak can be seen above sharply falling
backgrounds. Including the three bins nearest to $m_{jets}\sim 115$~GeV,
we obtain 4704 signal events and 1046 background events. Thus, the
signal-to-background ratio (S/B) is expected to be 4.5 after all cuts, and
the statistical precision in the signal rate measurement is expected
to be 2.3\%. If the Higgs factory is made with a 103~GeV electrons instead
of 75~GeV the S/B = 2.5 is not so favorable, because of the broader 
$\gamma\gamma$
energy distribution (see Table~\ref{tab:lum} and Fig.~\ref{fig:spectra}).

%--------------------------------------
\noindent\underline{$H \rightarrow WW $:}\\
%~\\
Observation of this decay mode is extremely difficult at high-energy
$\gamma\gamma$ colliders, because of the large cross section for $W$~pair
production.  If the $\gamma\gamma$ center-of-mass energy is below the
$W^+W^-$ threshold, however, the continuum production of $W$ pairs is
greatly reduced, allowing the observation of resonant production through a
Higgs boson.  The sharp peak in the $\gamma\gamma$ luminosity function seen in
Fig.~\ref{fig:spectra} plays a key role here.
Figure~\ref{fig:wwcross}(a) compares the cross sections for the 
continuum $W$~pair production with the Higgs resonance curve. 
%%%%%%%%%%%%%%%%%%%%%%%%%%%%%%%%%%%%%%%%%%%%%%%%%%%%%%%%%%%%%%%%%%%
%%%%%%%%%%%%%%%% F I G U R E %%%%%%%%%%%%%%%%%%%%%%%%%%%%%%%%%%%%%%%%%%%%%%%%%%
%%%%%%%%%%%%%%%%%%%%%%%%%%%%%%%%%%%%%%%%%%%%%%%%%%%%%%%%%%%%%%%%%%%%%%%%%%%%%%%
\begin{figure}[tbp]
\begin{center}
\resizebox{\textwidth}{!}
{\epsfig{file=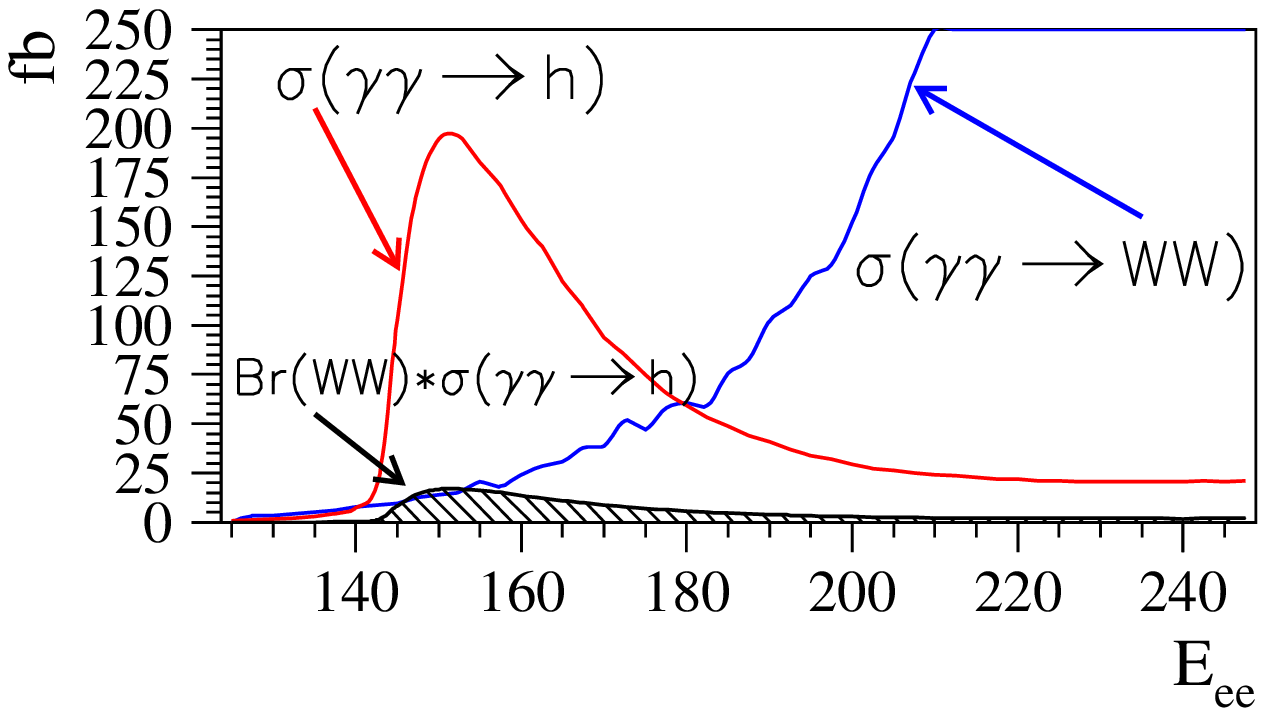,height=6.5cm}
\epsfig{file=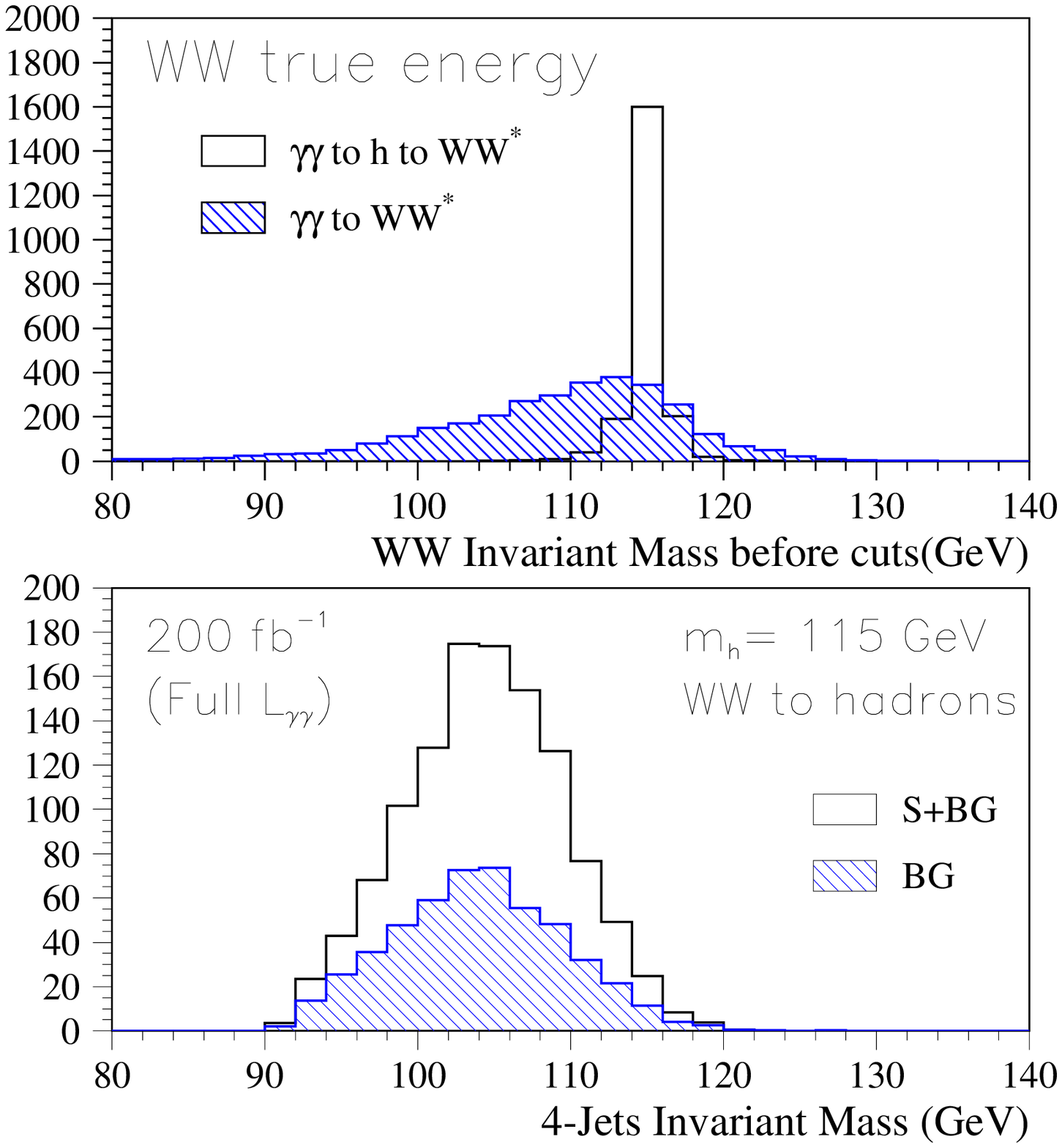,height=10.5cm}}
\end{center}
\caption[.]{\label{fig:wwcross}      
\em
(a) Cross sections for $\gamma\gamma\rightarrow h$, 
$\gamma\gamma\rightarrow h \times {\cal B}r(h\to WW)$ for 
$m_H=115$~GeV and  $\gamma\gamma\rightarrow WW$
production. (b) Comparison of the  ideal invariant mass
of the $WW$ pairs  from signal and background events.
(c) Selection of the $WW$   decay mode of the Higgs boson for 
$m_H=115$~GeV, running at $\sqrt{s_{\gamma\gamma}}=115$~GeV at CLICHE.
}
\end{figure}
%%%%%%%%%%%%%%%%%%%%%%%%%%%%%%%%%%%%%%%%%%%%%%%%%%%%%%%%%%%%%%%%%%%
As shown,
the cross sections for $\sigma(\gamma\gamma\to W^+W^-)$ and 
${\cal B}r(H\to W^+W^-) \times \sigma(\gamma\gamma\to H)$ are comparable,
if  $E_{CM}(e^-e^-)=150$~GeV for a  $m_H=115$~GeV. 
One significant difference between the two type of events is
the energy distribution of the $W^+W^-$ pairs, as illustrated 
in Fig.~\ref{fig:wwcross}(b).

Our study is concentrated on the hadronic decays of the $W$ pairs 
as described in ~\cite{cliche}.
After all   cuts we have a 29\% reconstruction efficiency.
A comparison of the signal and the  background after cuts is given in
Fig.~\ref{fig:wwcross}(c), which corresponds to a signal-to-background ratio
of 1.3, and the statistical precision in the
signal rate measurement is expected to be 5\%.

%--------------------------------------
\noindent\underline{$H \rightarrow \gamma\gamma $:}\\
%~\\
The decay $H \rightarrow
\gamma\gamma$ is  very rare. However,
the number of Higgs events is large at a $\gamma\gamma$~collider, 
so an interesting number of $H \to \gamma\gamma$
events would be produced.  Furthermore, the backgrounds are expected
to be quite small, below 2~fb~\cite{jikia_gg}, 
 since there is no tree-level coupling of photons,
and the box-mediated processes for $\ggc\rightarrow\ggc$  are peaked very sharply in the
forward direction. Initial estimates indicate that a clear peak
in the $\gamma\gamma$~mass distribution should be observable.

The number of events produced in this channel is proportional
to ${\Gamma_{\gamma\gamma}^2/\Gamma_{\mathrm{total}}}$. The quadratic
dependence is interesting, because if $\Gamma_{\mathrm{total}}$ could
be measured elsewhere, a small error on $\Gamma_{\gamma\gamma}$ would
be obtained.  In Fig.~\ref{fig:ggtogg}, we can see that a 8\% measurement of
${\Gamma_{\gamma\gamma}^2/\Gamma_{Total}}$ can be made with an integrated
luminosity of 175~fb$^{-1}$ or 40~fb$^{-1}$ at the  ${\cal L}_{\ggc}^{peak}$ 
at CLICHE.  From the comparison in Table~\ref{tab:lum}, we can see that 
this requires around    one year of data taking at CLICHE, and almost two times
longer for NLC.
%%%%%%%%%%%%%%%%%%%%%%%%%%%%%%%%%%%%%%%%%%%%%%%%%%%%%%%%%%%%%%%%%%%
%%%%%%%%%%%%%%%% F I G U R E %%%%%%%%%%%%%%%%%%%%%%%%%%%%%%%%%%%%%%%%%%%%%%%%%%
%%%%%%%%%%%%%%%%%%%%%%%%%%%%%%%%%%%%%%%%%%%%%%%%%%%%%%%%%%%%%%%%%%%%%%%%%%%%%%%
\begin{figure}[tbp]
\begin{center}
\mbox{\epsfig{file=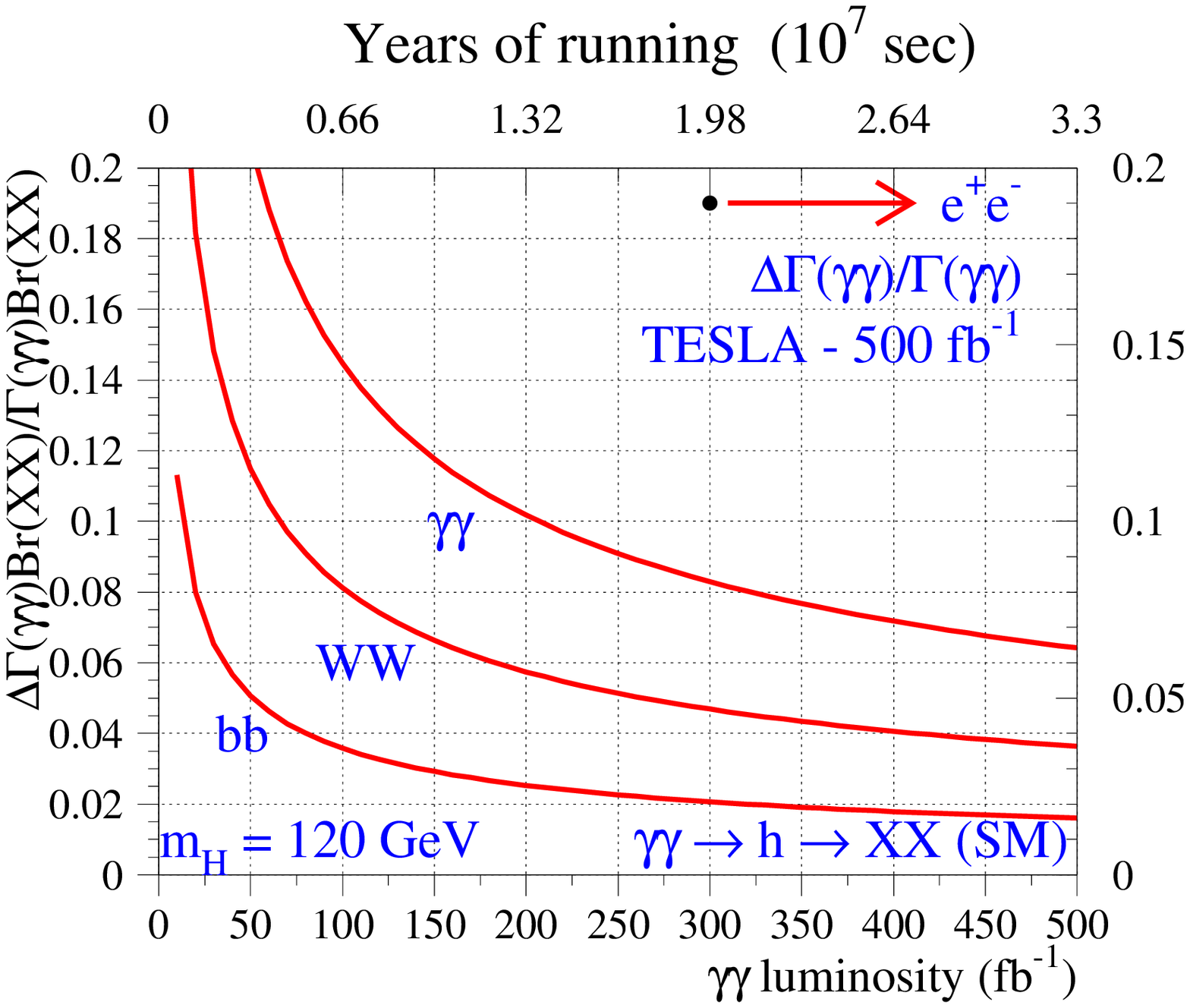,height=7.cm}}
\mbox{\epsfig{file=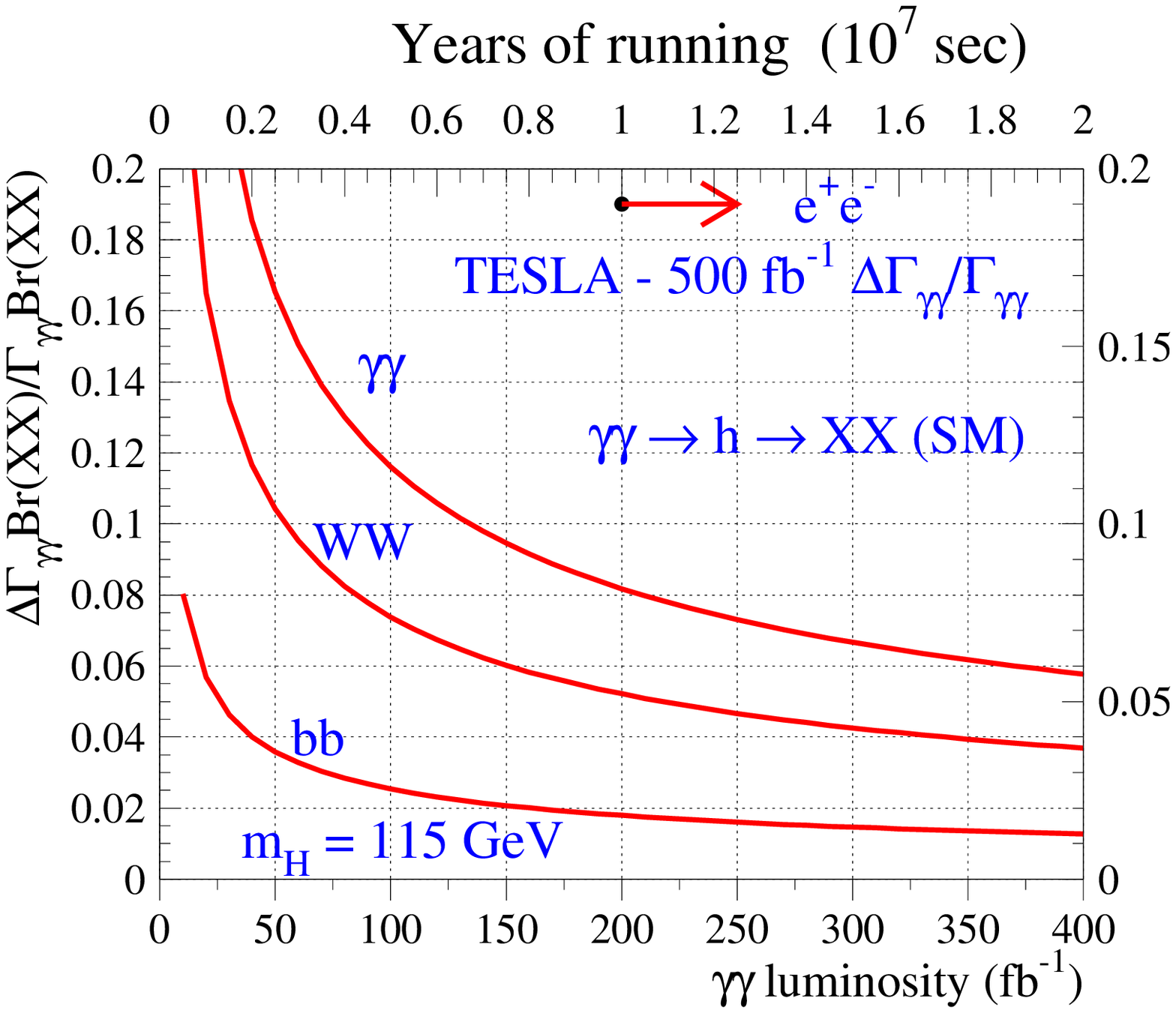,height=7.cm}}
\end{center}
\caption[.]{\label{fig:ggtogg}      
\em
The expected precision in the ${H \rightarrow \gamma\gamma}$ decay
width from direct measurements
of $H\rightarrow\gamma\gamma$ for $m_H = 120$~GeV at NLC with 
103~GeV electrons, and for  $m_H = 115$~GeV at CLICHE with 75~GeV electrons. 
The precision is
less than in the equivalent measurement of $H\rightarrow WW, \bar{b}b$,
but this observable is unique to a $\gamma\gamma$ collider.
}
\end{figure}
%%%%%%%%%%%%%%%%%%%%%%%%%%%%%%%%%%%%%%%%%%%%%%%%%%%%%%%%%%%%%%%%%%%

%%%%%%%%%%%%%%%%%%%%%%%%%%%%%%%%%%%%%%%%%%%%%%%%%%%%%%%%%%%%%%%%%%%%%%%%%%%%%%

\subsection{Using Higgs factory measurements: the MSSM}

\noindent\underline{$h \to b \bar b$ and $h \to WW^*$:}\\
Taking the ratio of rates
($\gamma \gamma \to h \to b \bar b$)/($\gamma \gamma \to h \to W W^*$),
the production 
cross section, total Higgs width and luminosity uncertainty
cancel, yielding the ratio of Higgs partial widths 
$\Gamma(h \to b \bar b)/\Gamma(h \to WW^*)$
with about 6\% statistical precision.
In the MSSM, this ratio can deviate from its Standard Model (SM) value.

The SM ratio
$\Gamma(H_{\rm SM} \to b \bar b)/\Gamma(H_{\rm SM} \to WW^*)$
depends strongly on the Higgs mass, varying by three orders of magnitude
over the range $100 \, {\rm GeV} < m_H < 160$ GeV.
The Higgs mass measurement at the Photon Collider with an 
uncertainty of 100 MeV 
%LHC if $h \to \gamma \gamma$ is close to its SM rate, 
yields a 1\% uncertainty in 
$\Gamma(H_{\rm SM} \to b \bar b)/\Gamma(H_{\rm SM} \to WW^*)$.  
This uncertainty is small compared to the expected experimental 
uncertainty.
An additional theoretical uncertainty
of 3.5\% in $\Gamma(H_{\rm SM} \to b \bar b)$ 
is due to the uncertainties in the $b$ quark mass and in $\alpha_s$ 
\cite{CHLM}.

In the MSSM, $\Gamma(h \to b \bar b)/\Gamma(h \to WW^*)$ 
generally differs from its SM prediction,
except in the decoupling limit \cite{decoupling}.  
For large pseudoscalar Higgs mass $m_A$ and 
$\tan\beta$ greater than a few,
%\begin{equation}
%	\frac{\Gamma(h \to b \bar b)/\Gamma(H_{\rm SM} \to b \bar b)}
%	{\Gamma(h \to WW^*)/\Gamma(H_{\rm SM} \to WW^*)}
%	= \left( 1 - \tan\beta \frac{\cos(\beta-\alpha)}{\sin(\beta-\alpha)}
%	\left[1 - \frac{\Delta_b}{1 + \Delta_b} \frac{1}{\sin^2\beta} \right]
%	\right)^2,
%	\label{eq:bW1}
%\end{equation}
\begin{equation}
	\frac{\Gamma(h \to b \bar b)/\Gamma(H_{\rm SM} \to b \bar b)}
	{\Gamma(h \to WW^*)/\Gamma(H_{\rm SM} \to WW^*)}
	\simeq 1 + \frac{4cm_Z^2}{m_A^2} 
	\left[1 - \frac{\Delta_b}{1 + \Delta_b} \right]
	+ \mathcal{O}(m_Z^4/m_A^4),
	\label{eq:bW2}
\end{equation}
where $c$ parameterizes the radiative corrections to the Higgs mixing
angle $\alpha$ (see Ref.~\cite{CHLM} for details)
and $\Delta_b$ is a $\tan\beta$-enhanced SUSY correction to the relation 
between the $b$ quark mass
and its Yukawa coupling~\cite{Deltab,COPW}.
%\begin{equation}
%	\Delta_b \simeq \left[ 
%        \frac{2 \alpha_s}{3 \pi} \mu M_{\tilde g} \,
%        I(M^2_{\tilde b_1}, M^2_{\tilde b_2}, M^2_{\tilde g})
%        + \frac{h_t^2}{16 \pi^2} \mu A_t \,
%        I(M^2_{\tilde t_1}, M^2_{\tilde t_2}, \mu^2)\right]\tan\beta\,,
%	\label{eq:Deltab}
%\end{equation}
%where $I(a,b,c)$ is given in Ref.~\cite{COPW}.  Note the $\tan\beta$ 
%enhancement.
Figure~\ref{fig:bWMSSM}(a) shows
$\Gamma(h \to b \bar b)/\Gamma(h \to WW^*)$ in the MSSM 
normalized to its SM value as a function of $m_A$.
%%%%%%%%%%%%%%%%%%%%%%%%%%%%%%%%%%%%%%%%%%%%%%%%%%%%%%%%%%%%%%%%%%%
%%%%%%%%%%%%%%%% F I G U R E %%%%%%%%%%%%%%%%%%%%%%%%%%%%%%%%%%%%%%%%%%%%%%%%%%
%%%%%%%%%%%%%%%%%%%%%%%%%%%%%%%%%%%%%%%%%%%%%%%%%%%%%%%%%%%%%%%%%%%%%%%%%%%%%%%
\begin{figure}[tbp]
\resizebox{\textwidth}{!}
{\rotatebox{270}{\includegraphics{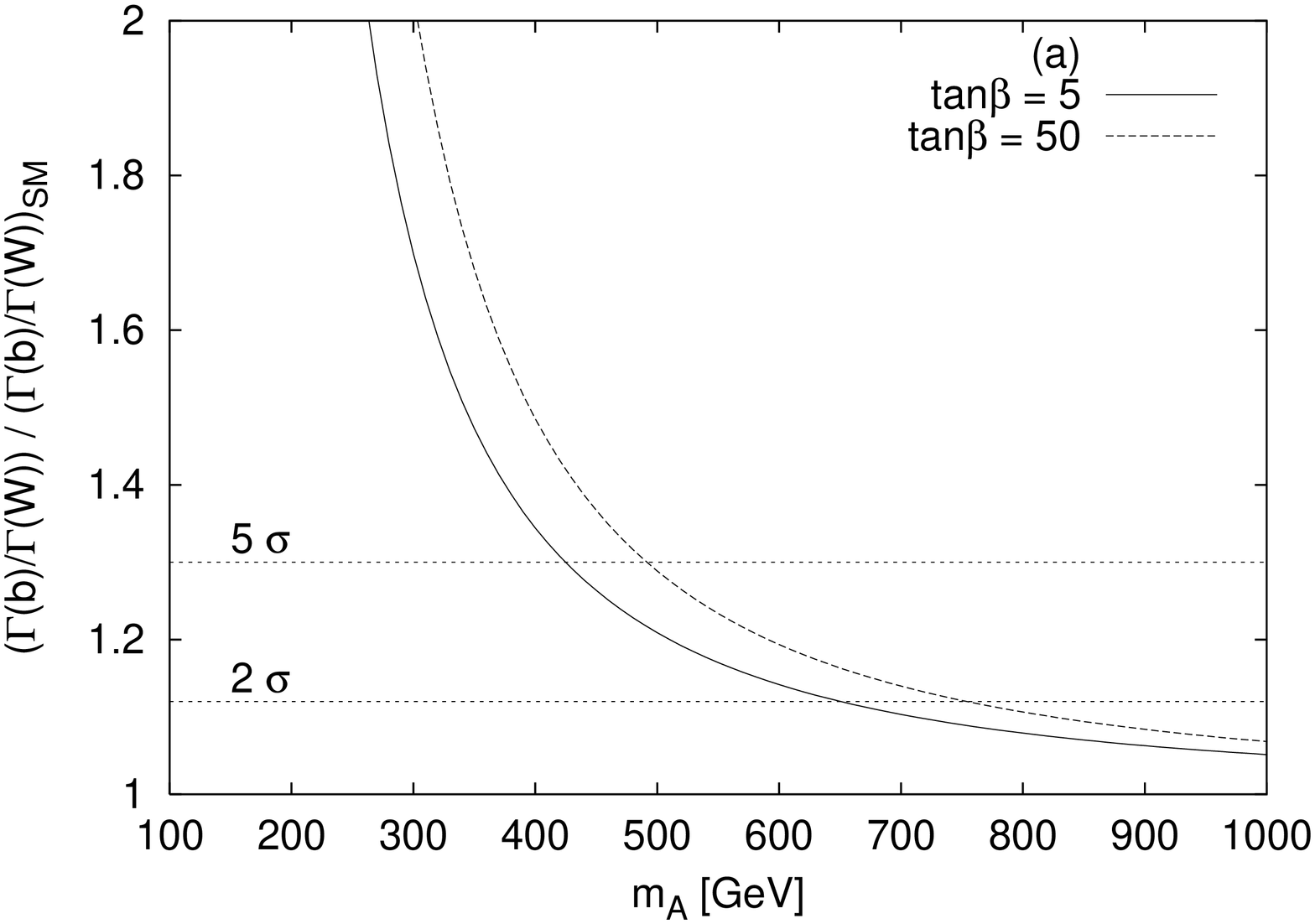}}
\rotatebox{270}{\includegraphics{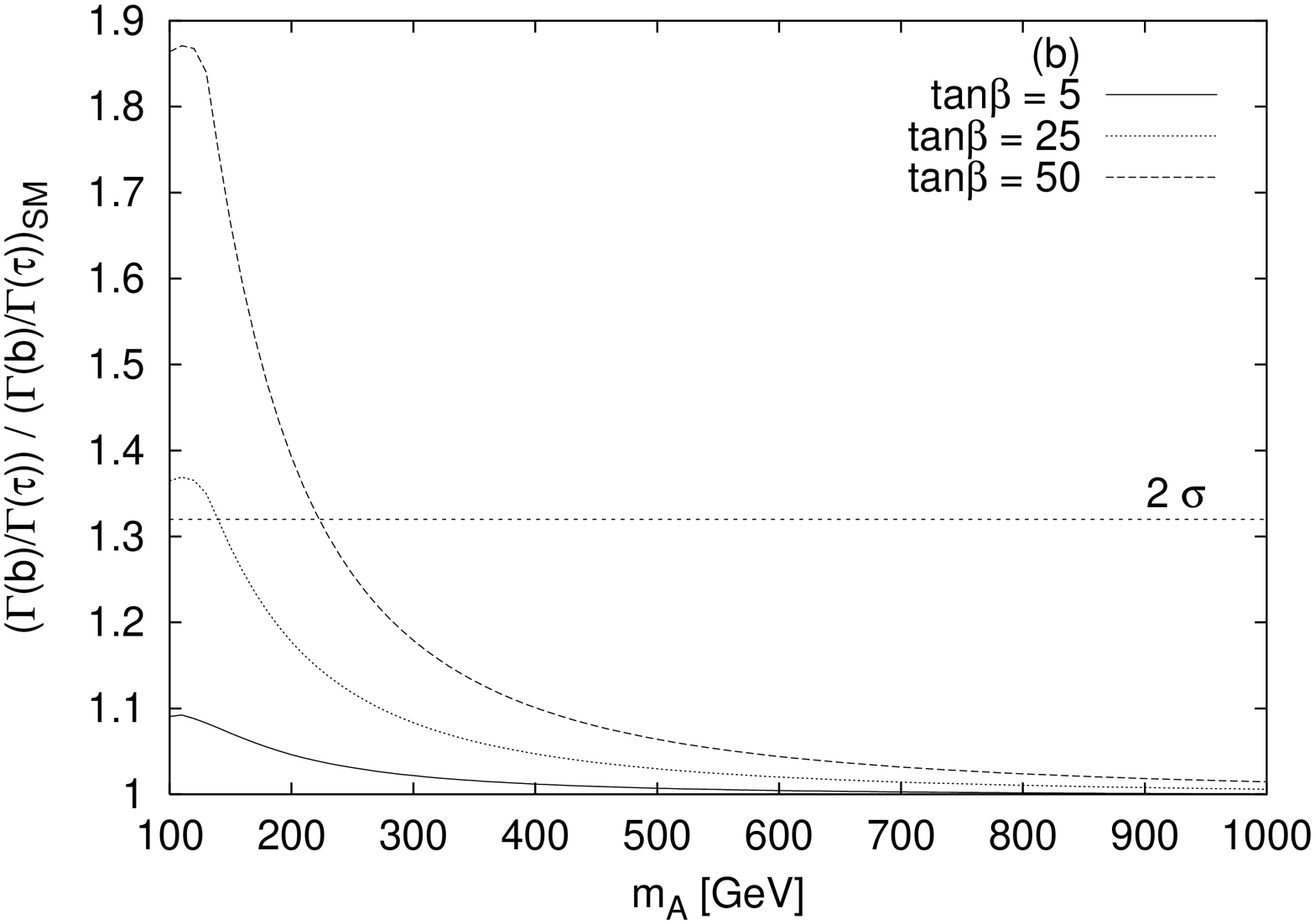}}}
\caption{\em (a) $\Gamma(h \to b \bar b)/\Gamma(h \to WW^*)$ in the MSSM
normalized to its SM value. We exhibit
2$\sigma$ and 5$\sigma$ deviations based on a 
6\% measurement.
(b) $\Gamma(h \to b \bar b)/\Gamma(h \to \tau \tau)$ 
in the MSSM normalized to its SM value.  
The 2$\sigma$ deviation shown is based on a 16\% measurement.
The MSSM parameters are $M_{\tilde Q} = M_{\tilde U} = M_{\tilde g} = 1$ TeV, 
$M_2 = 2M_1 = -\mu = 200$ GeV, 
$X_t \equiv A_t - \mu \cot\beta = \sqrt{6} M_{\tilde Q}$,
and $A_b = A_t$ (i.e., a maximal mixing scenario).  Numerical results
are obtained from HDECAY \cite{hdecay} with the $\Delta_b$ corrections added.}
\label{fig:bWMSSM}
\end{figure}
%%%%%%%%%%%%%%%%%%%%%%%%%%%%%%%%%%%%%%%%%%%%%%%%%%%%%%%%%%%%%%%%%%%%%%%%%%%%%%%
For the chosen MSSM parameters, a 6\% measurement of 
$\Gamma(h \to b \bar b)/\Gamma(h \to WW^*)$ will reveal a discrepancy 
from the SM at the 5$\sigma$ (2$\sigma$) level 
for $m_A \lsim 400$ GeV (650 GeV).

The reach in $m_A$ quoted above holds whenever the factor $c$ in 
(\ref{eq:bW2}) is close to one, as it is over most of the MSSM
parameter space.
However, there are small parameter regions in which $c$ is close to zero,
leading to decoupling even at low values of $m_A$ \cite{CHLM}.  
In such regions, $\Gamma(h \to b \bar b)/\Gamma(h \to WW^*)$ is very close 
to its SM value even for low $m_A$, and $\gamma \gamma$ collider 
measurements may
not reveal a discrepancy from the SM.

\noindent\underline{$h \to b \bar b$ and $h \to \tau^+\tau^-$:}\\
At the CERN Large Hadron Collider (LHC) the ratio
$\Gamma(h \to \tau \tau)/\Gamma(h \to WW^*)$ can be 
measured with a precision of about 15\% 
using Higgs production in weak boson fusion \cite{Zeppenfeld}.
Combining this measurement with $\Gamma(h \to b \bar b)/\Gamma(h \to WW^*)$ 
from the Photon Collider yields 
$\Gamma(h \to b \bar b)/\Gamma(h \to \tau \tau)$ 
with a precision of about 16\%.
This ratio is particularly sensitive to $\Delta_b$;
at large $\tan\beta$ and large $m_A$,
\begin{equation}
	\frac{\Gamma(h \to b \bar b)/\Gamma(H_{\rm SM} \to b \bar b)}
	{\Gamma(h \to \tau \tau)/\Gamma(H_{\rm SM} \to \tau \tau)}
	\simeq 1 - \frac{4cm_Z^2}{m_A^2} \frac{\Delta_b}{1 + \Delta_b}
	+ \mathcal{O}(m_Z^4/m_A^4).
\end{equation}
Figure~\ref{fig:bWMSSM}(b) shows
$\Gamma(h \to b \bar b)/\Gamma(h \to \tau \tau)$ in the MSSM normalized to its 
SM value as a function of $m_A$.
The $\tan\beta$ dependence of $\Delta_b$ 
%(Eq.~\ref{eq:Deltab})
is clearly visible; at large $\tan\beta$ and $m_A \lsim 225$ GeV, 
a 16\% measurement of $\Gamma(h \to b \bar b)/\Gamma(h \to \tau \tau)$ 
will reveal a discrepancy from the SM at the 2$\sigma$ level for the
chosen MSSM parameters.

\noindent\underline{Event rates:}\\
The rates
$\sigma(\gamma \gamma \to h) \times {\rm BR}(h \to b \bar b)$
and $\sigma(\gamma \gamma \to h) \times {\rm BR}(h \to WW^*)$
are directly measured at the $\gamma \gamma$ collider.
In the MSSM, these rates are generally expected to deviate from 
their SM predictions.  This leads to additional sensitivity to the 
possible MSSM nature of a light Higgs boson, and 
a full analysis of these deviations should be performed.  
In particular, if the production cross section and/or either
of the decay rates are suppressed compared to their SM values, the 
statistical precision of the $\gamma \gamma$ collider measurements may 
suffer.

%%%%%%%%%%%%%%%%%%%%%%%%%%%%%%%%%%%%%%%%%%%%%%%%%%%%%%%%%%%%%%%%%%%%%%%%%
\subsection{CP violation in Higgs couplings}

CP violation in Higgs couplings has been previously considered 
at $\gamma\gamma$ colliders in~\cite{GunionCP,Asakawattbar};
these analyses require linearly polarized initial-state photons 
\cite{GunionCP} or interference of final-state fermions with the
continuum in $\gamma\gamma \to H \to t \bar t$ \cite{Asakawattbar}.
These analyses probe CP violation in the Higgs couplings to $\gamma\gamma$
and $t \bar t$ pairs, respectively.  As pointed out in~\cite{FNALLC},
CP violation measurements in Higgs production and decay probe both CP mixing
in the Higgs mass eigenstate and CP violation in the Higgs couplings to 
external particles.  Thus CP violation measurements in many different
Higgs couplings are desirable.

Here we consider observables that probe CP violation in the Higgs coupling
to $W$ boson pairs.  We make no assumptions about the photon polarization,
so that this study can be done with a $\gamma \gamma$ collider running as 
a Higgs
factory, as described before.
Our CP-odd observables are constructed in such a way
that they are directly measurable in experiment without
reconstruction
of the $W$ boson rest frames or the center-of-momentum frame
of the initial pair of photons.  Thus they can be measured using semileptonic
$W$ decays, despite the unknown momentum carried off by the neutrinos.

We consider the process $\gamma\gamma \to H \to W^+W^-$.
%, with amplitude
%\begin{equation}
% T_{fi}=\varepsilon_{\nu_1}(p_1)\varepsilon_{\nu_2}(p_2)
%      \varepsilon^*_{\mu_1}(k_1)\varepsilon^*_{\mu_2}(k_2)
%    A^{\nu_1\nu_2\mu_1\mu_2}(p_1,p_2,k_1,k_2).
%\end{equation}
We assume that the polarization of the initial photons is not known. 
Then, a CP test in this mode is possible only if the polarizations of 
the $W^+$ or $W^-$ are observed. 
To obtain information about the polarizations we consider
leptonic decays of the $W$ bosons:
\begin{equation}
 \gamma (p_1) +\gamma (p_2) \rightarrow H \rightarrow W^+(k_1)+W^-(k_2)
   \rightarrow \ell^+(q_1)+\ell^-(q_2) +{\rm neutrinos},
	\label{eq:ggHWW}
\end{equation}
where all the momenta are defined in the $\gamma\gamma$ c.m.\ frame.
The process $\gamma\gamma \to H \to W^+W^-$ proceeds with an amplitude
\begin{equation}
 	T_{fi}=\varepsilon_{\nu_1}(p_1)\varepsilon_{\nu_2}(p_2)
      	\varepsilon^*_{\mu_1}(k_1)\varepsilon^*_{\mu_2}(k_2)
   	A^{\nu_1\nu_2\mu_1\mu_2}(p_1,p_2,k_1,k_2).
\end{equation}
For the decay process $W^+(k_1)\rightarrow \ell^+(q_1)+\nu $ 
we define a covariant decay matrix $\rho^+_{\mu\nu}(k_1,q_1)$
%which is normalized as:
%\begin{equation}
%	\frac{1}{4\pi (k_1^0)^2}
%     	\int {d\Omega_1 \over (1-\beta {\bf \hat k_1}\cdot
% 	{\bf \hat q_1})^2}\rho^+_{\mu\nu}(k_1,q_1)
%	=\left(-g_{\mu\nu}+{k_{1\mu}k_{1\nu}\over M_W^2}\right)
%\end{equation}
%where ${\bf \hat k_1}={\bf k_1}/\vert {\bf k_1}\vert$,
%${\bf \hat q_1}={\bf q _1}/ \vert {\bf q_1}\vert$,
%$\beta=\vert {\bf k_1}\vert / k_1^0$,
%and $\Omega_1$ is the solid angle of ${\bf q_1}$. 
%At tree-level in the SM, the decay matrix
%$\rho^+_{\mu\nu}(k_1,q_1)$ takes the form:
%\begin{equation}
%	\rho^+_{\mu\nu}(k_1,q_1)={3\over 2}\left(-M_W^2g_{\mu\nu}
%   	+2(k_{1\mu}q_{1\nu}+k_{1\nu}q_{1\mu})-4q_{1\mu}q_{1\nu}
%    	+2i\varepsilon_{\mu\nu\alpha\beta}k_1^\alpha q_1^\beta\right),
%\end{equation}
and similarly for the $W^-$ decay matrix $\rho^-_{\mu\nu}(k_2,q_2)$.
The probability for the process in (\ref{eq:ggHWW})
may be written as:
\begin{equation}
	{ R({\bf p_1},{\bf k_1},{\bf q_1},{\bf q_2})=
     	{1\over 4}A^{\nu_1\nu_2\mu_1\mu_2}(p_1,p_2,k_1,k_2)
    	{ A^*_{\nu_1\nu_2}}^{\mu'_1\mu'_2}(p_1,p_2,k_1) \nonumber \\
    	\rho^+_{\mu_1\mu'_1}(k_1,q_1)\rho^-_{\mu_2\mu'_2}(k_2,q_2)}.
\end{equation}
If CP invariance holds, then $R({\bf p_1},{\bf k_1},{\bf q_1},{\bf q_2})=
R({\bf p_1},{\bf k_1},-{\bf q_2},-{\bf q_1})$.

The expectation value of any observable $O$ that is a function of
${\bf p_1},{\bf k_1},{\bf q_1}$ and ${\bf q_2}$ can be obtained from
\bea  
	\langle O \rangle 
	&=& {1\over N} \int f_\gamma (x_1)f_\gamma (x_2) {dx_1dx_2
  	\over x_1x_2}
	\frac{\beta}{2(4\pi k_1^0)^4} 
	\int d\Omega
     	{d\Omega_1 \over (1-\beta {\bf \hat k_1}\cdot
 	{\bf \hat q_1})^2} % \nonumber \\
     	%&& \times 
	{d\Omega_2\over (1+\beta {\bf \hat k_1}\cdot
 	{\bf \hat q_2})^2} 
  	O R({\bf p_1},{\bf k_1},{\bf q_1},{\bf q_2}),
	\label{eq:Oexp}
\eea
normalized so that $\langle 1 \rangle =1$.
Here, $d\Omega$, $d\Omega_1$ and $d\Omega_2$ are 
the solid angle of ${\bf k_1}$, ${\bf q_1}$ and ${\bf q_2}$, respectively,
and $f_\gamma (x)$ gives the proportion of the photons
with the fraction $x$ of the initial electron or 
positron beam energy.
Experimental cuts can be included in (\ref{eq:Oexp}), but for our
purpose they must be CP blind.

The momenta ${\bf p_1}$, ${\bf k_1}$, ${\bf q_1}$ and ${\bf q_2}$ are not
directly measurable due to the missing neutrinos and
the lack of knowledge about the $\gamma\gamma$ c.m.\ frame.
To construct CP-odd observables, we use the lepton
momenta, which are directly measured in experiment and are related to 
${\bf q_1}$
and ${\bf q_2}$ through a Lorentz boost. We denote these momenta by
$q_+=(E_+,{\bf q_+})$ and $q_-=(E_-, {\bf q_-})$ for
$\ell^+$ and $\ell^-$, respectively.
We construct the following CP odd observables:
\bea
 O_1 = {E_+-E_- \over M_W}, \qquad
                O_2 = ({\bf \hat p}\cdot {\bf \hat q_+})^2
                     -({\bf \hat p}\cdot {\bf \hat q_-})^2, \qquad
                O_3 = {\bf \hat p}\cdot ({\bf \hat q_+}-{\bf\hat q_-})
                      {\bf\hat p}\cdot ({\bf\hat q_+}\times {\bf\hat q_-}),
\eea
with
${\bf\hat q_+} ={\bf q_+}/\vert {\bf q_+}\vert$,
${\bf\hat q_-} ={\bf q_-}/\vert {\bf q_-}\vert$,
and the vector ${\bf \hat p}$ is the direction of motion of the electron or
positron.
Because of the Bose symmetry of the two-photon initial state, the 
expectation
value of any observable which is odd in ${\bf\hat p}$ is zero. With these
observables one can also define the corresponding CP asymmetries:
\begin{equation}
   	A_i={ N(O_i >0) -N(O_i <0) \over N(O_i>0)+N(O_i<0)}, \qquad (i=1,2,3)
\end{equation}
where $N(O_i >0)$ ($N(O_i <0)$) denotes the number of events with $O_i >0$ 
($O_i <0$).
Any nonzero $<O_i>$ or any nonzero $A_i$ indicates CP violation.
Further, the observables
$O_1$ and $O_2$ are CPT odd, the expectation values of them and the
corresponding asymmetries can be nonzero only if an absorptive part of the
amplitude and CP violation exist. 

Summarizing, CP-odd observables and the
corresponding CP asymmetries 
for the process $\gamma\gamma \to H \to WW$ can be
constructed from the directly measured
energies and momenta of the leptons from the $W$ decays. 
With these observables one can detect CP
violation without requiring complete knowledge of the center-of-momentum
frame of the initial photons or of the rest frame of the $W$ bosons.
Therefore, our observables are easy to measure on an event-by-event basis.
An estimate gives a statistical error of $\delta A/A\sim 5\%$ or better  
from the measurements 
that could be made at CLICHE, see section~\ref{l_higgs}.

%%%%%%%%%%%%%%%%%%%%%%%%%%%%%%%%%%%%%%%%%%%%%%%%%%%%%%%%%%%%%%%%%%%%%%%%%
%\section{Physics opportunities -- Heavy Higgs Bosons}
\subsection{The Heavy MSSM Higgs Bosons $\hh,\ha$}

In many scenarios, it is likely that we will
observe small deviations from SM expectations in precision measurements
of the properties of the SM-like Higgs boson, 
and thus suspect the presence of heavy Higgs bosons. 
However, direct production of the heavier Higgs bosons in $\epem$ 
collisions is likely to require large machine energy. 
For example, in the MSSM, $\epem\to \hh\ha$ 
is the most relevant process in the decoupling limit, but
requires $\rts>\mhh+\mha$, with $\mhh+\mha\sim 2\mha$ as the decoupling
limit sets in. 
The alternative modes $e^+e^- \to b\anti b \hh$, $b\anti b \ha$
are only viable if $\tanb$ is large~\cite{Grzadkowski:2000wj}. 
At the LHC, either low or high $\tanb$ is
required for discovery of $\hh,\ha$ if they have mass $\gsim 250\gev$,
as seen in Fig.~\ref{f:atlasmssm}.  
\begin{figure}[tbp]
\centerline{\resizebox{85mm}{!}
	{\psfig{file=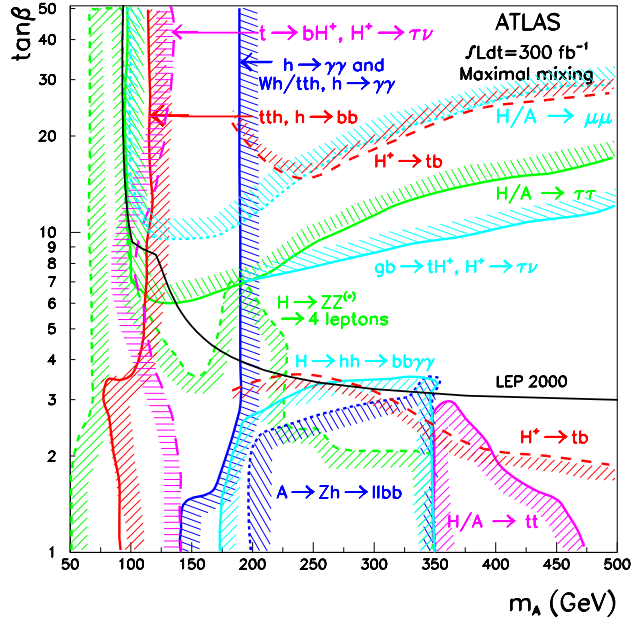,width=14cm}}}
\caption[0]{\em $5\sigma$ discovery contours for MSSM Higgs boson detection
at the LHC in various channels,
assuming maximal mixing and an integrated luminosity of $L=300\fbi$
for the ATLAS detector. This figure is preliminary \cite{atlasmaxmix}.}
  \label{f:atlasmssm}
\end{figure}
After accumulation of $L=300\fbi$ at the LHC, 
$\hh,\ha$ will be detected except in the wedge of parameter
space with $\mha\gsim 250\gev$ and moderate $\tanb$, where only
$\hl$ can be detected. If the $e^+e^-$ LC is operated at $\rts=630\gev$,
then detection of $\epem\to\hh\ha$ will be possible for $\mha\sim\mhh$
up to nearly 300 GeV. 
In this case, some other means of detecting $\hh,\ha$ must be
found in the portion of the LHC wedge with $\mha\gsim 300\gev$.

We show here that single $\hh,\ha$ production via $\gam\gam$
collisions will allow their discovery throughout a large fraction
of this wedge: see~\cite{GunionHA} for details. 
The event rate can be substantial due to quark loop contributions ($t$
and, at high $\tanb$, $b$) and loops containing SUSY
particles.
In this study, we assume that the
masses of the superparticles (charginos, squarks, sleptons, \etc) 
are sufficiently heavy that $\hh,\ha$ do not decay to them
and that the superparticle loop contributions to the $\gam\gam(\hh,\ha)$
couplings are negligible.

If we have no reliable prior constraints on $\mha,\mhh$, 
an important question is whether it is best to search for the $\hh,\ha$
by scanning in $\rts$ (and thereby in $E_{\gam\gam}$)
using a peaked spectrum,
or running at fixed $\rts$
using a broad $E_{\gam\gam}$ spectrum part of the time and a peaked 
spectrum the rest of the time \cite{Gunion:1993ce}.
Our results indicate that if covering the LHC wedge
region is the goal, then running at a single
energy, half the time with a peaked $E_{\gam\gam}$ luminosity distribution
and half the time with a broad distribution,
is likely to be the optimal approach.

\begin{figure}[tbp]
\centerline{\resizebox{130mm}{!}{\psfig{file=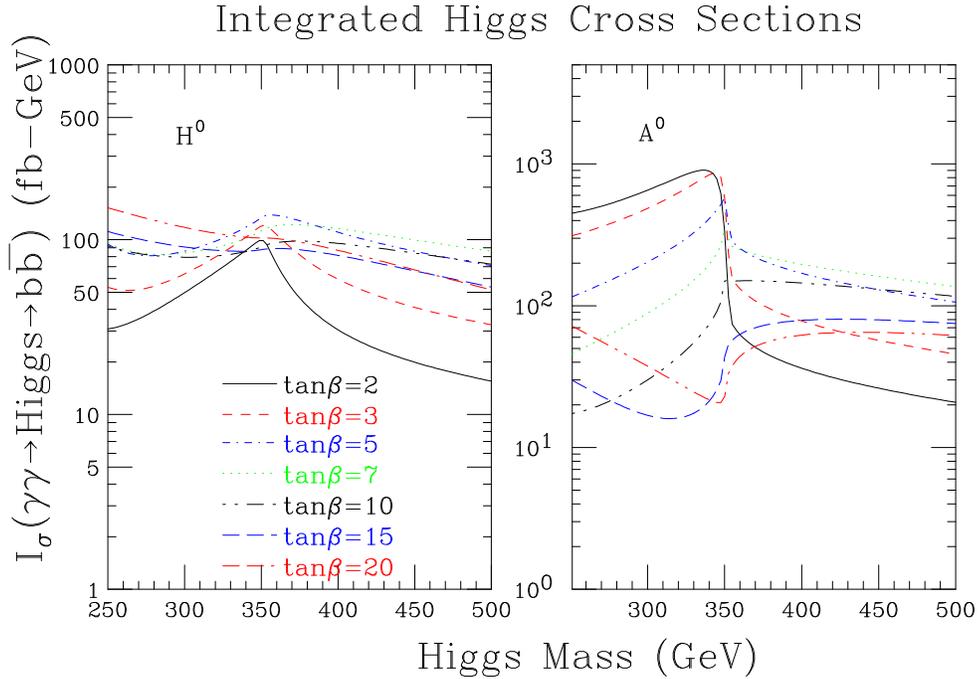,height=12cm}}}
\caption[0]{\em Integrated $\hh$ and $\ha$ Higgs cross sections
as defined in~\cite{GunionHA} as a function of $\mha$, for a variety
of $\tanb$ values. We assume the maximal-mixing scenario with
$\msusy=1\tev$. Supersymmetric particle loops are neglected.  
}
\label{f:sigeff}
\end{figure}
The effective integrated cross sections for 
$\gamma\gamma \to \hh,\ha \to b \bar b$,
taking into account acceptance and cuts, are 
plotted as a function of $\mha$ 
for a variety of $\tanb$ values in Fig.~\ref{f:sigeff} 
(see~\cite{GunionHA} for details).  Numerical results were obtained
using HDECAY \cite{hdecay}, with $\mt=175\gev$, 
$\msusy=1\tev$ for all slepton and squark soft-SUSY-breaking
masses and $\mu=+1\tev$.  We also assume the maximal-mixing
scenario, $A_t=\mu/\tanb+\sqrt 6 \msusy$, and $A_b=A_\tau=A_t$.
If the LC is operated at $\rts=630\gev$ (corresponding to $x\sim 5.69$
for 1 micron laser wavelength) we can potentially probe Higgs masses up to
$\sim 500\gev$.

The photon energy and polarization spectra are computed using the
CAIN \cite{cain2} Monte Carlo.  For the broad spectrum,
the luminosity remains quite large even below the $E_{\gam\gam}$
peak at $E_{\gam\gam}=500\gev$, and the polarization combination 
$\vev{\lam\lam'}$ is large for $E_{\gam\gam}>450\gev$. 
For the peaked spectrum,
the luminosity is substantial for $E_{\gam\gam}=400\gev$
and rises rapidly with decreasing $E_{\gam\gam}$. In addition,
reasonably large $\vev{\lam\lam'}$ is retained for $250<E_{\gam\gam}<400\gev$.
However, in both cases, $1-\vev{\lam\lam'}$ is always large enough
that the $J_z=2$ part of the $b\anti b$ background will be dominant.
In order to detect the Higgs bosons with mass substantially below
the machine energy of 630 GeV, we must employ cuts that remove
as little luminosity for $E_{\gam\gam}$ substantially below $\rts$ 
as possible while still eliminating most of the 
$b\anti b(g)$ and $c\anti c (g)$ backgrounds.
A cut on $|\cos\theta^*|<0.5$ (where $\theta^*$
is the angle of the $b$ jets in the $\gam\gam$ rest frame) 
eliminates much of the ($t$-channel) background while decreasing
the ($s$-channel) signal by only a factor of two.
A second cut is imposed upon the $m_{b\anti b}$ mass distribution.
The optimal value for this cut depends upon the Higgs widths,
the degree of degeneracy of the $\hh$ and $\ha$ masses,
and the detector resolutions and reconstruction techniques. 
For the $\tanb$ range inside the problematical wedge ($15>\tanb>3$), the
$\ha$ and $\hh$ are still relatively narrow, 
with widths below $1-2$ GeV.
Thus, the width of the $b\anti b$ mass distribution derives mostly
from detector resolutions and reconstruction procedures.
A full Monte Carlo analysis for heavy 
Higgs bosons with relatively small widths is not yet available.
However, there are many claims in the literature that the resulting
mass resolution will almost certainly be better than 
$30\%/\sqrt{m_{b\anti b}}$ (the result obtained assuming $18\%/\sqrt{E_{jet}}$
for each of the $b$ 
jets)~\cite{nlc_report,Abe:2001gc,TESLA_TDR}.
Very roughly, this corresponds to a full-width at half maximum of
about $6\gev$ in the mass range from $250-500\gev$ of interest.
We adopt the procedure of considering a 10 GeV bin centered on the
Higgs mass in question and assume that 50\% of the Higgs events will fall
into this bin.  This would be very conservative for the 6 GeV full-width
estimate {\em assuming that the $\hh$ and $\ha$ are degenerate in mass.}
In practice, they are not exactly degenerate and so we have used the 10 GeV
as a conservative approach to allowing for this non-degeneracy.

\begin{figure}[tbp]%[h!]
\begin{center}
\resizebox{\textwidth}{60mm}
{\epsfig{file=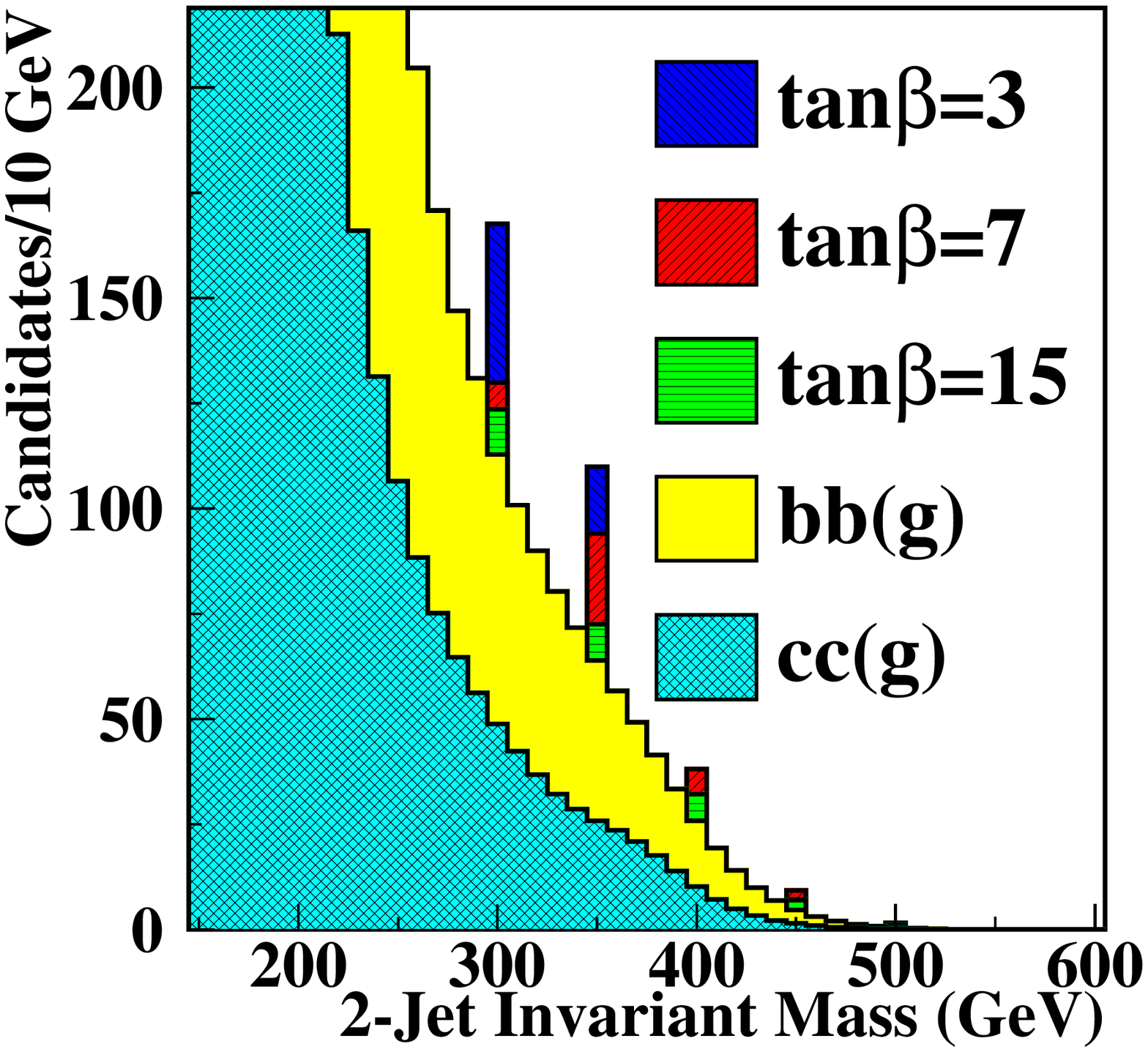,height=4.5cm}
\epsfig{file=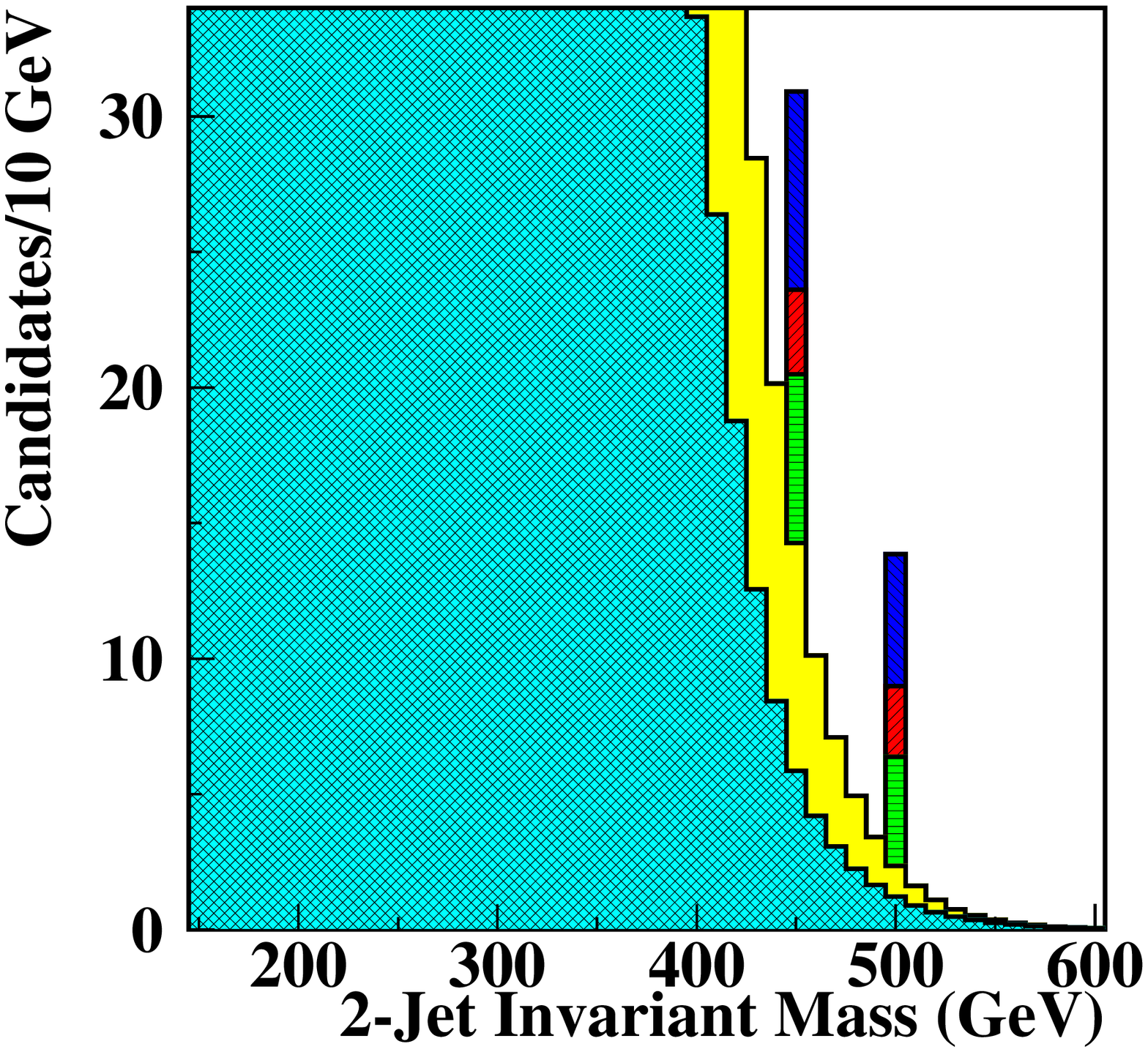,height=4.5cm}}
\vspace*{-.5in}
\end{center}
\caption[0]{\em
Signal and background rates for the
$[\mha,\tanb]$ cases considered for
(a) broad spectrum and (b) peaked spectrum 
operation at $\rts=630\gev$ for one year ($10^7$ sec). 
The signals shown assume that
50\% of the total number of signal events fall into the single 10 GeV bin
shown. Signals in the side bins are not shown.
}
\label{signalploti}
\end{figure}
The resulting signals and backgrounds after cuts are shown in 
Fig.~\ref{signalploti} as a function of 2-jet invariant mass with the
signals superimposed (we plot only the central 10 GeV bin assumed 
to contain 50\% of the signal events).  
Results for different values of $\tanb$ and $\mha$ are shown
for running in the broad and peaked spectra configurations.
Note that for the $\mha=350\gev$ points, we have conservatively run HDECAY
so that $\mhh,\mha$ are slightly above the $t \bar t$ threshold.
As can be seen from Fig.~\ref{f:sigeff}, the rates (especially
that for $\gam\gam\to \ha\to b\anti b$) depend sensitively 
on the Higgs masses relative to the $t \bar t$ threshold.
For $\mha$ just below the plotted 350 GeV points, 
the net signal is much stronger.

Many of the $[\mha,\tanb]$ cases considered will yield an observable 
$4\sigma$ signal.
Our ability to cover the LHC wedge in which the neutral $\hh,\ha$ Higgs 
bosons cannot be detected is illustrated in Fig.~\ref{wedgeplot}.
(The $\hpm$ can be detected at the LHC down to lower $\tanb$ values
than can the $\hh,\ha$, as shown in Fig.~\ref{f:atlasmssm}.)
\begin{figure}[tbp]%[t!]
\begin{center}
\resizebox{\textwidth}{100mm}
{\epsfig{file=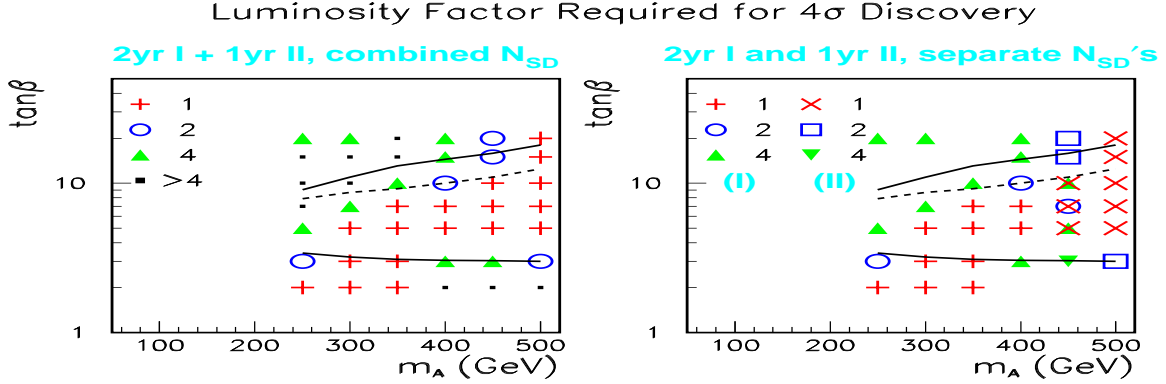}}
\end{center}
%\vspace*{-3.7in}
\vspace*{-2.1in}
\caption[0]{\em 
The $[\mha,\tanb]$ points for which two years of broad spectrum operation 
plus one year of peaked spectrum operation at $\rts=630\gev$
will yield $S/\sqrt{B}\geq4$.  
Shown are (a) the combined significance from both the broad spectrum 
and peaked spectrum running
and (b) the separate significances from the broad spectrum and peaked
spectrum running.
Also shown are the additional points for
which a $4\sigma$ signal is achieved if the total
luminosity is doubled (`2') or quadrupled (`4')
relative to our assumed luminosity.
Such luminosity
increases could be achieved by improved technical designs and/or longer run
times; {\it e.g.}, the luminosity expected from the TESLA design
corresponds roughly to the doubled luminosity (`2').
The small black squares in (a) indicate additional
points sampled for which even a luminosity increase by a factor
of four for both spectra does not yield a $4\sigma$ signal.
The solid curves show the boundaries of the LHC wedge region from
Fig.~\ref{f:atlasmssm} --- the lower black curve is that from the
LEP (maximal-mixing) limits, but is
somewhat higher than that currently claimed by the LEP electroweak working 
group.
For $\tan\beta$ values above the dashed curve,
$\hpm\to\tau^\pm\nu_\tau$ can be directly detected
at the LHC. 
}
\label{wedgeplot}
\end{figure}
After running for two years in the broad spectrum configuration,
7 of the 13 $[\mha,\tanb]$ cases considered in the LHC wedge region 
with $\mha=300,350,400\gev$ will yield a $4\sigma$ or greater Higgs signal,
with the best sensitivity at low to moderate $\tan\beta$.
Similarly, after running for one year in the peaked spectrum configuration,
7 of the 10 $[\mha,\tanb]$ cases considered in the LHC wedge with 
$\mha=450,500\gev$ will yield a $4\sigma$ or greater Higgs signal,
with the best sensitivity at moderate to high $\tan\beta$.
These points are shown in Fig.~\ref{wedgeplot}(b).
The areas of parameter space covered by the broad spectrum and the 
peaked spectrum running are complementary;
if data from both broad spectrum and peaked spectrum running are 
combined for a given parameter space point, the statistical significance
is only slightly improved.  
In all, a $4\sigma$ or greater Higgs signal would be detected for 15 of the 23
points considered in the LHC wedge, a coverage of about 65\%.
Further improvements in luminosity or mass resolution would
be helpful for guaranteeing complete coverage of the wedge region.
If the luminosity is doubled for both the broad spectrum and peaked spectrum 
running, the coverage increases to 78\%.
In addition, for $\rts=630\gev$ it is very probable that
one could see $e^+e^- \to \hh\ha$ pair production 
for $\mha=300\gev$, in which case $\gam\gam$ operation with doubled luminosity
would allow detection of $\hh,\ha$ 
throughout most of the remaining portion of the wedge in which they cannot
be seen by other means (see Fig.~\ref{wedgeplot}(b)).
We also note that in this study we have considered only $b\bar b$ final
states.  At low $\tanb$, we expect that 
the $\hh \to \hl\hl$ and $\ha \to Z\hl$ channels 
will provide observable signals for the remaining points
with $\mha\leq 2\mt=350\gev$ in the LHC wedge. 
The $t\anti t$ channels might provide
further confirmation for $b\anti b$ signals for wedge points with 
$\mha>450\gev$.  
Finally, we note that the single most difficult wedge point considered 
is $\mha=400,\tanb=15$, which is at the edge of the LHC wedge.
The region of the LHC wedge in which our running scenario would not
enable $\hh,\ha$ detection in $\gam\gam$ collisions is roughly given by
$325\gev\lsim\mha\lsim 400\gev$ and $\tanb>8$.  In this region, though,
the LHC would be able to detect the charged Higgs boson via
$\hpm\to \tau^\pm\nu_\tau$ (see Fig.~\ref{f:atlasmssm}) and measure 
its mass to about $\pm 25\gev$.
If studies of the sparticles
indicate that the MSSM is the correct theory, then 
we would expect $\mha\sim\mhh\sim \mhpm$,
and could then run the $\gam\gam$ collider with a peaked 
spectrum at the $\rts$ value yielding $E_{\rm peak}\sim \mhpm$.

A rough determination of $\tanb$ is likely
to be possible using the data associated with the initial discovery
of $\hh,\ha$ in $\gamma\gamma$ collisions.
We show the approximate fractional error on $\tan\beta$ from the 
initial discovery data for the $[\mha,\tanb]$ points studied in
Table~\ref{tanberrors}.
Although the errors are not small, this
determination can be fruitfully combined with other $\tanb$ determinations,
especially for larger $\tanb$ values where other techniques
for determining $\tanb$ also have substantial errors.
More importantly, these results show clearly that a dedicated
measurement of the $\gam\gam\to \hh,\ha\to b\anti b$ rate 
and the rates in other channels 
($\hh\to\hl\hl$, $\ha\to Z\hl$, $\hh,\ha\to t\anti t$)
using a peaked
spectrum with $E_{\rm peak}=\mha$
is likely to yield a rather high precision determination of $\tanb$
after several years of optimized operation, and may provide information
about other supersymmetry parameters.
\begin{table}[ht]
%\large
\renewcommand{\arraystretch}{0.7}
  \begin{center}
\begin{tabular}[c]{|c|c|c|c|c|c|c|}
\hline $\mha(\gev)$ & 250 & 300 & 350 & 400 & 450 & 500 \\
\hline
$\tanb=2$ 
 & 0.51
 & 0.34
 & 0.20
 & 0.66
 & 0.46
 & 0.48
  \\
$\tanb=3$ 
 & 0.51
 & 0.27
 & $-$
 & 0.45
 & 0.30
 & 0.32
 \\
$\tanb=5$ 
 & 0.71
 & 0.34
 & 0.19
 & $-$
 & 0.56
 & 0.55
 \\
$\tanb=7$ 
 & $-$
 & 0.66
 & 0.23
 & 0.62
 & 0.67
 & 0.87
 \\
$\tanb=10$ 
 & $-$
 & $-$
 & 0.50
 & 0.64
 & 0.46
 & 0.53
 \\ 
$\tanb=15$ 
&  0.46
&  0.67
&  $-$
&  $-$
&  $-$
&  $-$
\\
\hline
    \end{tabular}
    \caption{\em Approximate uncertainties in $\tanb$ as determined
from measurements of the $\gam\gam\to \hh,\ha\to b\anti b$ rate
associated with Higgs discovery in the LHC wedge. These 
errors assume two years of operation in broad spectrum mode  
and one year of operation in peaked spectrum mode at $\rts=630\gev$.
Errors larger than 100\% are not shown.
}
    \label{tanberrors}
  \end{center}
\end{table}

%%%%%%%%%%%%%%%%%%%%%%%%%%%%%%%%%%%%%%%%%%%%%%%%%%%%%%%%%%%%%%%%%%%%%%%%%
\section{$e\gamma$ collider option -- doubly charged Higgs bosons}

Doubly charged Higgs bosons would have a distinct experimental
signature. Such particles arise in many extensions of 
the Standard Model (SM) such as the 
Higgs triplet model of Gelmini and Roncadelli~\cite{GRtrip} and the
left-right symmetric model.  
The signals for doubly charged Higgs bosons
arising from an $SU(2)_L$ triplet were studied in the 
process $e^-\gamma\to e^+ \mu^-\mu^-$.  
Details of the analysis are given in reference \cite{us} and 
contribution P3-18 of these proceedings.
The photon was assumed to be produced by backscattering a laser from the
$e^+$ beam of an $e^+e^-$ collider \cite{backlaser}.
We consider $e^+e^-$ center of mass 
energies of $\sqrt{s}=500$, 800, 1000, and 1500~GeV
appropriate to the TESLA/NLC/JLC high energy colliders 
and $\sqrt{s}=3$, 5, and 8~TeV for the CLIC proposal. 
In all cases an integrated luminosity of ${\cal L}=500$~fb$^{-1}$ was 
assumed. Because the signature of same sign 
muon pairs in the final state is so distinctive, with no SM background, 
the process can be sensitive to virtual $\Delta^{--}$'s with masses in
excess of the center of mass energy, depending on the strength of the
Yukawa coupling to leptons.  

Indirect
constraints on $\Delta$ masses and couplings have been obtained from lepton
number violating processes
\cite{swartz_and_other}. Rare  decay measurements  
\cite{mu3e_and_other} yield  very stringent restrictions on the
non-diagonal couplings $h_{e\mu}$ which were consequently neglected.
Stringent limits on flavor diagonal couplings 
come from the muonium anti-muonium conversion
measurement \cite{muoniumex} which 
requires that the ratio of the Yukawa coupling, $h$, and Higgs mass, 
$M_\Delta$, satisfy
$h/M_\Delta < 0.44$~TeV$^{-1}$ at 90\% C.L..
These bounds allow the existence of low-mass doubly charged Higgs with 
a small coupling constant.  
Direct search strategies for the $\Delta^{--}$ have been explored for 
hadron colliders \cite{datta_and_others}, with the mass reach at the LHC 
extending to $\sim 850$~GeV.
Signatures have also been explored for various configurations of
lepton colliders, including $e\gamma$ colliders. 

In the process $e^-\gamma \to e^+ \mu^-\mu^-$, the signal of 
like-sign muons is distinct and SM background free, 
offering excellent potential for doubly charged Higgs discovery. 
The process proceeds via the production of a positron along 
with a $\Delta^{--}$, with the subsequent $\Delta$ decay into two muons 
as well as through additional non-resonant contributions. 
%The cross section is a convolution of the backscattered 
%laser photon spectrum, $f_{\gamma/e}(x)$ \cite{backlaser}, 
%with the subprocess 
%cross section, $\hat{\sigma}(e^- \gamma \to e^+ \mu^-\mu^-)$.
Due to contributions to the final state that proceed 
via s-channel $\Delta^{--}$'s, the doubly-charged Higgs 
boson width must be included.  
Because the $\Delta$ width is model dependent, we account for the possible 
variation in width without restricting ourselves to 
specific scenarios by calculating the width using
$\Gamma (\Delta^{--}) = \Gamma_b + \Gamma_f$
where $\Gamma_b$ is the partial width to final state bosons and 
$\Gamma_f$ is the partial width into final state fermions.  
Two scenarios for the bosonic width were considered: 
a narrow width scenario with 
$\Gamma_b=1.5$~GeV and a broad width scenario with $\Gamma_b=10$~GeV. 
These choices represent a reasonable range for various values of the 
masses of the different Higgs bosons.
The partial width to final state fermions is given by
$\Gamma (\Delta^{--}\to \ell^- \ell^-) = \frac{1}{8\pi} 
h^2_{ \ell \ell} M_\Delta$.
Since we assume $h_{ ee} =h_{\mu\mu} =h_{\tau\tau} \equiv h$,
we have 
$\Gamma_f = 3 \times \Gamma (\Delta^{--}\to \ell^- \ell^-) $. 
%Many
%studies assume the $\Delta$ decay is entirely into leptons; for small
%values of the Yukawa coupling and relatively low $M_{\Delta}$ this leads
%to a width which is considerably narrower than our assumptions for 
%the partial width into bosons. 

We consider two possibilities for the $\Delta^{--}$ signal.  We assume
that either  all three final state particles are
observed and identified or that  the
positron is not observed, having been lost down the beam pipe.  
To take into account detector acceptance we restrict 
the angles of the observed particles relative to the beam, 
$\theta_{\mu},\; \theta_{e^+}$, to the ranges $|\cos \theta| \leq 0.9$.
We restrict the particle energies 
$E_{\mu}$, $E_{e^+} \geq 10$~GeV and assumed an identification 
efficiency for each of the detected final state particles of $\epsilon = 0.9$.  

Given that the signal for doubly charged Higgs bosons is so distinctive and 
SM background free, discovery would be signalled by even one event.
Because the value of the cross section for the process we consider is
rather sensitive to the $\Delta$ width, the potential for discovery 
of the $\Delta$ is likewise sensitive to this model dependent 
parameter.  Varying $\Gamma_b$, we find that, relative 
to $\Gamma_b = 10$~GeV, the case of zero bosonic width has a sensitivity 
to the Yukawa coupling $h$ which is greater by a factor of about 5
\cite{us}.

In Fig.~\ref{fig:Delta--}
we show 95\% probability (3 event) contours in the $h-M_{\Delta}$ 
parameter space. In each case, we assume the narrow width 
$\Gamma=1.5+\Gamma_f$~GeV case.  
Figure 1a corresponds to the center 
of mass energies  $\sqrt{s}=500$, 
800, 1000, and 1500~GeV,  
for the case of three observed particles in the
final state, whereas Fig. 1b shows the case where only the two muons are
observed.
Figs. 1c and 1d correspond to the energies being 
considered for the CLIC $e^+ e^-$ collider, namely, 
$\sqrt{s}=3$, 5, and 8~TeV,  for the three body and two body
final states, respectively.
In each case, for $\sqrt{s}$ above the $\Delta$ production threshold, 
the process is sensitive to the existence of the $\Delta^{--}$ with 
relatively small Yukawa couplings.  However, when the  $M_\Delta$ 
becomes too massive to be produced the values of the Yukawa couplings 
which would allow discovery grow larger slowly.
%%%%%%%%%%%%%%%%%%%%%%%%%%%%%%%%%%%%%%%%%%%%%%%%%%%%%%%%%%%%%%%%%%%
%%%%%%%%%%%%%%%% F I G U R E %%%%%%%%%%%%%%%%%%%%%%%%%%%%%%%%%%%%%%%%%%%%%%%%%%
%%%%%%%%%%%%%%%%%%%%%%%%%%%%%%%%%%%%%%%%%%%%%%%%%%%%%%%%%%%%%%%%%%%%%%%%%%%%%%%
\begin{figure}[t]
\centerline{
\begin{minipage}[t]{6.0cm}
%\begin{figure}[htbp]
\vspace*{13pt}
\centerline{
             \hspace{-0.6cm}
\includegraphics[width=5.8cm, height=6cm,angle=-90]{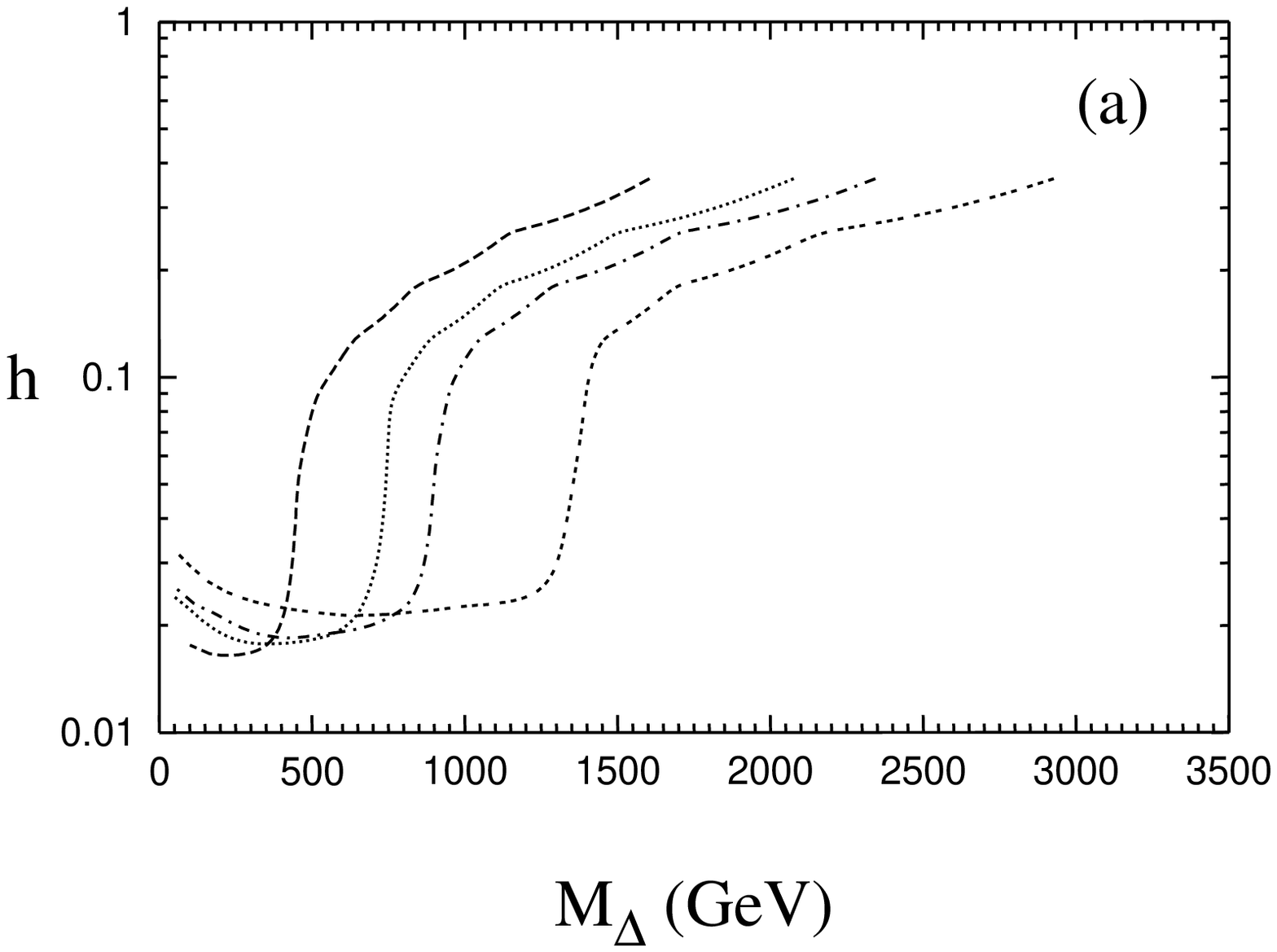}}%,width=6.1cm,clip}}
%\centerline{(6a)}
\vspace*{13pt}
\end{minipage} 
\hspace*{0.5cm}
\begin{minipage}[t]{6.0cm}
\vspace*{13pt}
\includegraphics[width=5.8cm, height=6.0cm,angle=-90]{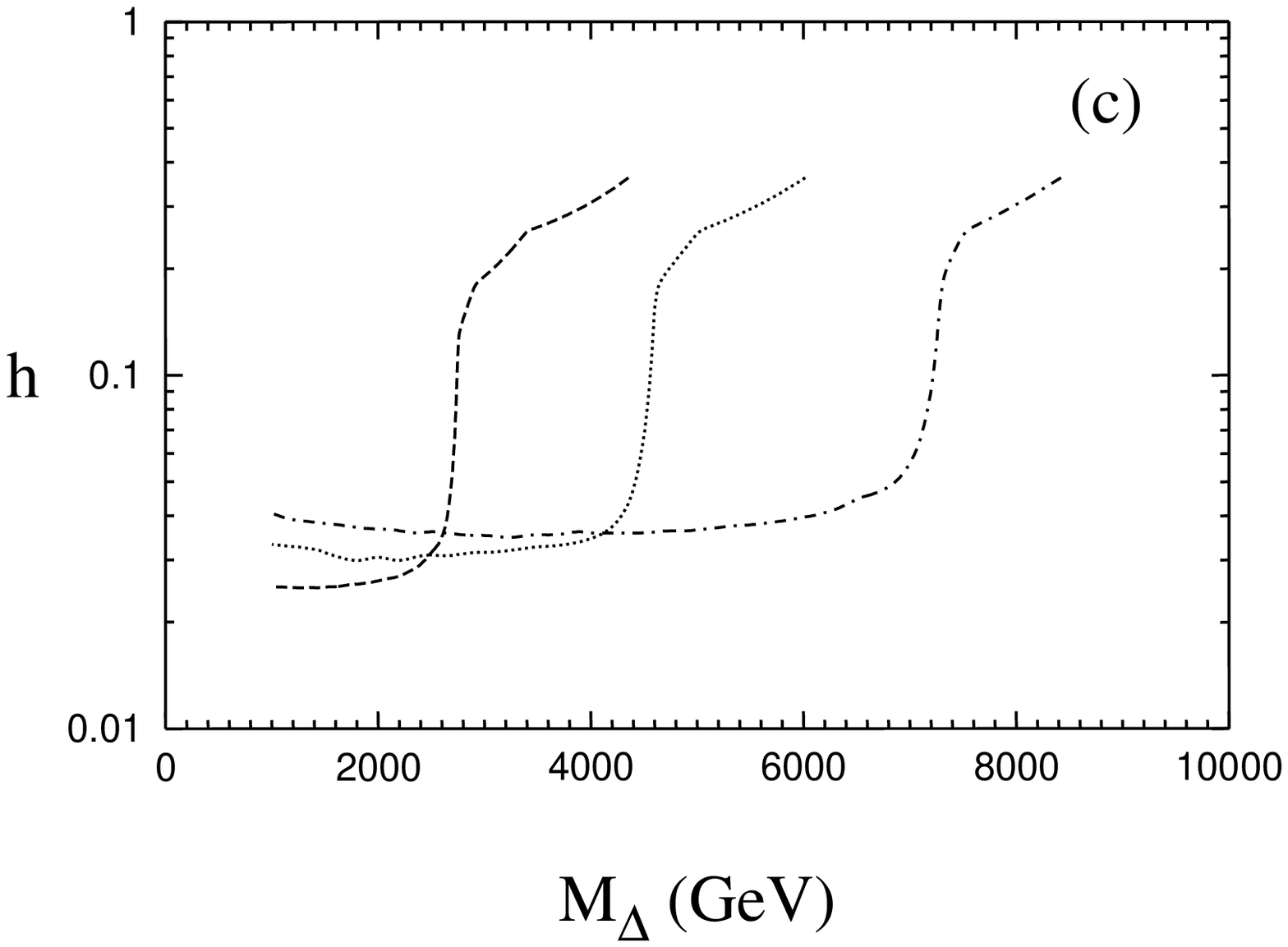}
\vspace*{13pt}
\end{minipage}
}   
\centerline{\hspace*{-0.6cm}
%\newline
\hspace*{0.5cm}
\begin{minipage}[t]{6.0cm}
\vspace*{13pt}
\includegraphics[width=5.8cm, height=6.0cm,angle=-90]{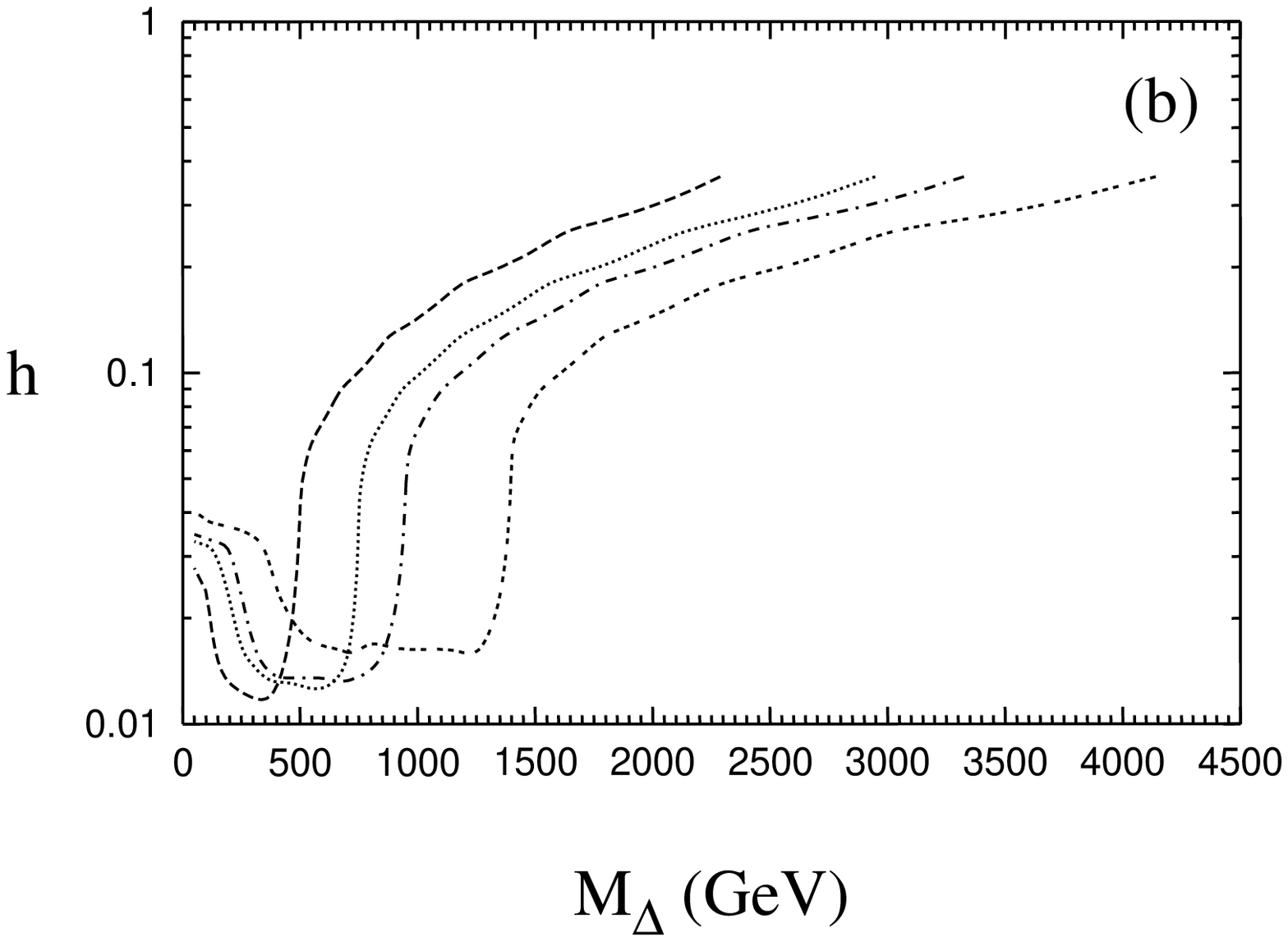}
%\centerline{(6b)}
\vspace*{13pt}
\end{minipage} 
\hspace*{0.5cm}
\begin{minipage}[t]{6.0cm}
\vspace*{13pt}
\includegraphics[width=5.8cm, height=6.0cm,angle=-90]{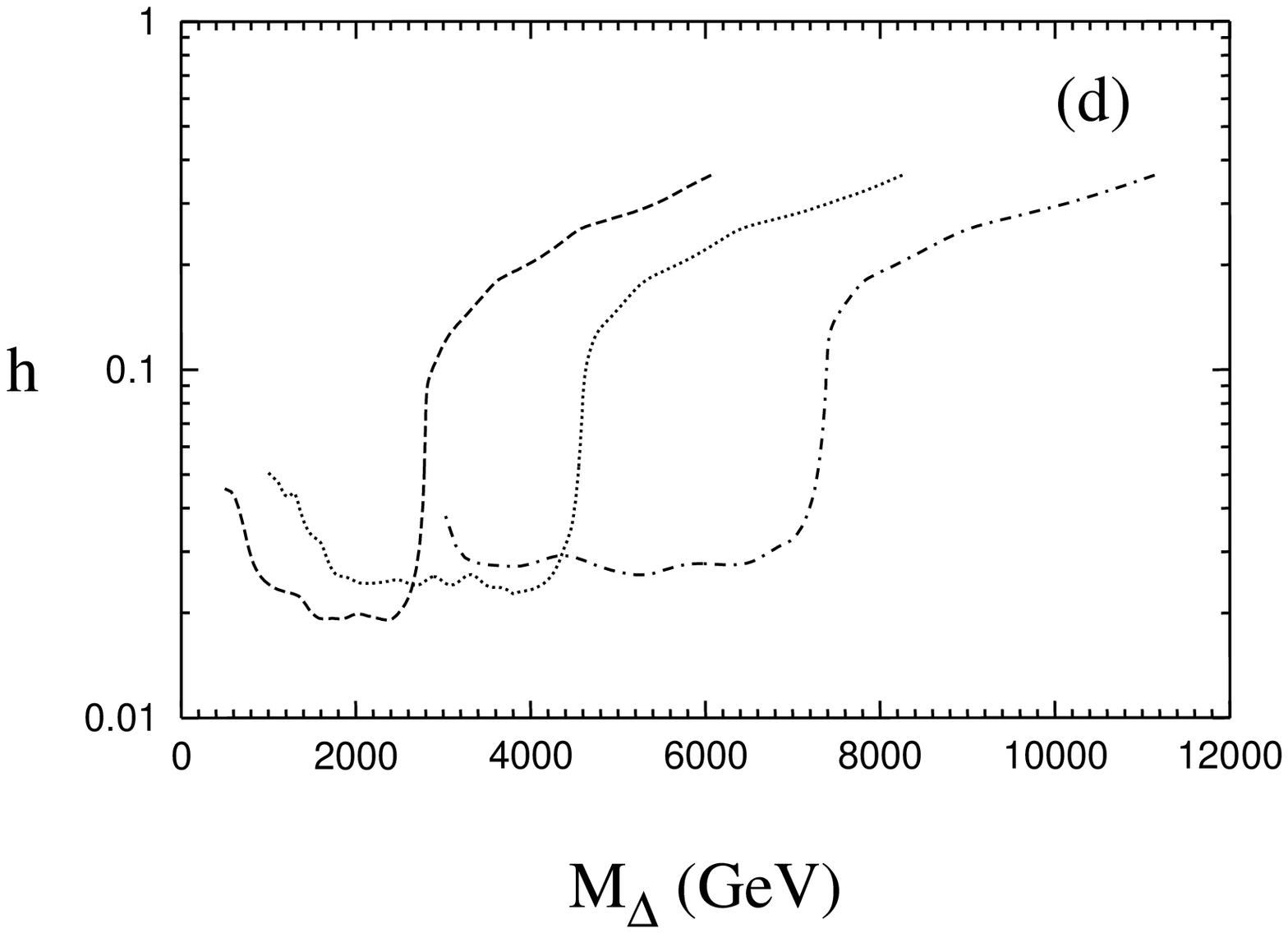}
%\centerline{(7b)}
\vspace*{13pt}
\end{minipage}
         }
\caption{ \em Discovery limits for the charged Higgs bosons
as a function of Yukawa coupling and $M_{\Delta}$.
(a) and (b) show TESLA/NLC/JLC center 
of mass energies $\sqrt{s}=500$, 800, 1000, and 1500~GeV,  
for the three particle and two particle final states, respectively.
(c) and (d) show CLIC center of mass energies
$\sqrt{s}=3$, 5, and 8~TeV, for the three particle and two particle
final states, respectively.
}
\label{fig:Delta--}
\end{figure}
%%%%%%%%%%%%%%%%%%%%%%%%%%%%%%%%%%%%%%%%%%%%%%%%%%%%%%%%%%%%%%%%%%%%%%%%%%%%%%%

The observation of doubly charged Higgs bosons would represent physics
beyond the SM and, as such, searches for this type of particle
should be part of the 
experimental program of any new high energy facility.  
We found that for $\sqrt{s_{e\gamma}}> M_\Delta$ doubly 
charged Higgs bosons could be discovered for even relatively small 
values of the Yukawa couplings; $h > 0.01$. For larger values of the 
Yukawa coupling the $\Delta$ should be produced in sufficient quantity 
to study its properties.   For values of $M_\Delta$ greater than the 
production threshold, discovery is still possible due to the 
distinctive, background free final 
state in the process  $e\gamma \to e^+ \mu^-\mu^-$ which can proceed 
via virtual contributions from intermediate $\Delta$'s.  Thus, even an 
$e^+e^-$ linear collider with modest energy has the potential to 
extend $\Delta$ search limits significantly higher than can be 
achieved at the LHC.

\section{Conclusions}

Our working group devoted most of its effort to exploring the various 
ways in which a $\gam\gam$ collider could contribute to our understanding
of Higgs physics.

For a SM-like Higgs boson, it will
be possible to determine $\Gamma(\gam\gam\to \h)\br(\h\to b\anti b), 
\Gamma(\gam\gam\to \h)\br(\h\to WW)$ and $\Gamma(\gam\gam\to \h)\br(\h\to \gam\gam)$
with excellent precision, \eg\ $\sim 2$, 5 and 8\%, respectively, 
for a  $\mh\sim 115\gev$.
In addition, the Higgs mass can be measured three ways (fitting the peaks
in the $\bar{b}b$ and $\gamma\gamma$ mass distributions, and by the
threshold method), and the partial width $\Gamma_{\gamma\gamma}$ can be
extracted on the basis of a measurement of ${\cal
B}r(H\rightarrow\bar{b}b)$ from an $e^+e^-$ machine to very good accuracy,
not matched by any other method.
At this level of accuracy, deviations that might be present
as the result of the SM-like Higgs boson being part of a larger Higgs
sector, such as that of the MSSM, would be visible if some
of the other Higgs bosons were not too much heavier than $500\gev$ or so.
The $WW$ decay mode will allow us to make a 5\% measurement of asymmetries
that are sensitive to the CP of the Higgs. In addition, 
a determination of  the CP nature of any Higgs boson can also be 
observed by employing
transversely (linearly) polarized laser beam photons~\cite{teslatdr,GunionHA}. 

For the purpose of building a light Higgs factory, the optimal operating
conditons are found to be  when we operate at the peaked $E_{\gam\gam}$ 
spectrum, that is obtained with
lower electron beam energy ($E=75-80$~GeV) combined with a  frequency 
tripler to reduce the wavelength of the available high power 1~micron 
lasers.

For the higher energy $\gamma\gamma$ collider,
 we conclude that it will be possible to detect $\ha,\hh$
of the MSSM Higgs sector using just $b\anti b$ states in 
a large fraction of the wedge of moderate-$\tanb$
space beginning at $\mha\gsim 300\gev$ 
(the approximate upper reach of the $\epem\to\hh\ha$
pair production process for $\rts=630\gev$) up to
the $E_{\gam\gam}$ spectrum limit of about 500 GeV, by running 
for two years with a broad spectrum and one year with a peaked spectrum,
without lowering the energy below $\rts=630\gev$.
By also considering
$\hh\to \hl\hl$, $\ha\to Z\hl$ and $\hh,\ha\to t\anti t$ final states,
we estimate that somewhat more than 85\% of the wedge parameter
region with $\mha\lsim 500\gev$
would provide a detectable signal after a total of
two to three years of operation.
Further, at all of the higher $\tanb$ points in the wedge region
for which $\gam\gam$ collisions would not allow detection of the $\hh,\ha$,
detection of $\hpm\to\tau^\pm\nu_\tau$ would be possible at the LHC.
Then, using the MSSM prediction $\mha\sim\mhh\sim\mhpm$ for 
$\mha\gsim 200\gev$, one could optimize the search for $\hh,\ha$ at
the $\gam\gam$ collider by running with a peaked
luminosity spectrum with $E_{\rm peak}=\mhpm$.
Thus, by combining $\gam\gam$ collider operation at $\rts=630\gev$ 
with $\epem$ running and LHC searches for the MSSM
Higgs bosons, it would be essentially guaranteed that we
could detect all the neutral Higgs
bosons of the MSSM Higgs sector if they have mass $\lsim 500\gev$, 
whereas without the $\gam\gam$
collider one would detect only the $\hl$ at both the LC and LHC
in the LHC wedge for $\mha\gsim 300\gev$. 
One caveat to this very optimistic set of conclusions is that
if supersymmetric particles are light enough to be produced in $\hh,\ha$ 
decays,
they will alter the $\gamma\gamma \to \hh,\ha$ cross sections
and reduce the $\hh,\ha\to b\anti b$ branching ratios, especially at 
low $\tan\beta$. 
In short, if we detect
supersymmetric particles at the LHC and LC consistent with the MSSM structure
and find only the $\hl$ at the LHC and LC, $\gam\gam$ operation
focusing on Higgs discovery will be a high priority.

\newpage

%%%%%%%%%%%%%%%%%%%%%%%%%%%%%%%%%%%%%%%%%%%%%%%%%%%%%%%%%%%%%%%%%%%%%%%%%%%
%%%%%%%%%%%%%%%%%%%%%%%%%%%%%%%%%%%%%%%%%%%%%%%%%%%%%%%%%%%%%%%%%%%%%%%%%%%

\end{document}
%
% ****** End of file template.snowmass ******